\documentclass[aps,pre,twocolumn,showpacs,superscriptaddress]{revtex4-1}  
\usepackage{graphicx}
\usepackage{dcolumn}
\usepackage{bm}   
\usepackage{color}
\usepackage{amssymb}
\usepackage{amsmath}

\newcommand{\sech}{\normalfont\mbox{sech}\,}
\newcommand{\e}{\normalfont\mbox{e}\,}

\begin{document}

\title{Manipulating localized matter waves in multi-component Bose-Einstein condensates}

\author{K. Manikandan}
\affiliation{Centre for Nonlinear Dynamics, Bharathidasan University, Tiruchirappalli 620024, Tamilnadu, India}
\author{P. Muruganandam}
\affiliation{Department of Physics, Bharathidasan University, Tiruchirappalli 620024, Tamilnadu, India}
\author{M. Senthilvelan}
\affiliation{Centre for Nonlinear Dynamics, Bharathidasan University, Tiruchirappalli 620024, Tamilnadu, India}
\author{M. Lakshmanan}
\affiliation{Centre for Nonlinear Dynamics, Bharathidasan University, Tiruchirappalli 620024, Tamilnadu, India}

\begin{abstract}
We analyze vector localized solutions of two-component Bose-Einstein condensates (BECs) with variable nonlinearity parameter and external trap potential through similarity transformation technique which transforms the two coupled Gross-Pitaevskii equations into a pair of coupled nonlinear Schr\"{o}dinger equations with constant coefficients under a specific integrability condition.  In this analysis we consider three different types of external trap potentials: a time-independent trap, a time-dependent monotonic trap, and a time-dependent periodic trap.  We point out the existence of different interesting localized structures, namely rogue waves, dark-and bright soliton-rogue wave, and rogue wave-breather-like wave for the above three cases of trap potentials. We show how the vector localized density profiles in a constant background get deformed when we tune the strength of the trap parameter.  Further we investigate the nature of the trajectories of the nonautonomous rogue waves. We also construct the dark-dark rogue wave solution for repulsive-repulsive interaction of two-component BECs and analyze the associated characteristics for the three different kinds of traps.  We then deduce single, two and three composite rogue waves for three component BECs and discuss the correlated characteristics when we tune the strength of the trap parameter for different trap potentials.  
\end{abstract}
\maketitle

\section{Introduction}
The study of multi-component nonlinear waves is one of the fascinating topics which has potential applications in Bose-Einstein condensates (BECs) in atomic physics \cite{kevre} and optial fibers in nonlinear optics \cite{kivs}.  The realization of BECs in weakly interacting atomic gases has strongly stimulated a large number of studies on exploring nonlinear properties of matter waves such as envelope solitons, gap solitons, soliton chains and so on \cite{anderson,davis}.  The development in trapping techniques for BECs has allowed experimentalists to simulataneously confine atomic clouds in different hyperfine spin states or different atomic species.  The first experiment involving interaction between multiple-species BECs was performed in rubidium atoms which demonstrated the possibility of producing long-lived multiple condensate systems \cite{myatt}.  In particular, the experimental demonstration of trapped multi-species BECs have spurred great excitement in atomic physics and stimulated interest in studying various properties of two-component BECs \cite{peth:smi}. A mixture of BECs can be produced experimentally by simultaneously trapping atoms in different hyperfine states or two isotopes of the same element or of different species \cite{mathews}. Recent experimental observations have also shown that the dark-bright and dark-dark vector solitons can be formed in certain two-species trapped dilute-gas BECs \cite{hamner,hoefer}.  In contrast to the single-component case, such multicomponent condensates can present novel and fundamentally different scenarios for the collective dynamics and coherent structures due to the intercomponent interactions.

Rogue waves (RWs) are random nonlinear waves which occasionally rise up in the ocean which can attain amplitudes more than twice the value of background wave field \cite{osbrn,khar:pelin}.  They appear from nowhere and disappear without a trace \cite{osbrn:rato}.  RWs were also noticed in a variety of physical systems such as optical fibers \cite{solli}, superfluids \cite{ganshin} and capillary waves \cite{shatz}. It was observed that modulational instability \cite{benj:feir}  is responsible for the sudden rise in the wave amplitude in the ocean \cite{pere}. Recently, higher order rational solutions have been shown to attain even higher amplitudes of wave in the description of the RW phenomenon \cite{akmv:anki}.  On the other hand, there has been a number of studies committed to RWs in coupled nonlinear Schr\"{o}dinger equations (CNLSE).  The CNLSE also models water wave interactions \cite{zakha} and wave propagation in fiber communication system \cite{solli}.  The exact analytical RW solutions of the CNLSE (integrable Manakov case) \cite{fabio,vishnu} and its generalized version have been presented in \cite{vishnu1}. Further, BECs have also been shown to constitute a good platform to explore the RWs which allow one to understand deeply the nature and dynamics of RWs under laboratory conditions.  The possibility of identifying vector RWs in multi-component BECs provides us a rich phenomenology of nonlinear wave structures.

Serkin et al. have considered the nonautonomous scalar NLS equation with linear and harmonic oscillator potentials and explored many specific features of the non-autonomous solitons \cite{serkin}.  Followed by this study, some further interesting studies of the nonautonomous NLS models with external potentials have also been made \cite{atre,raj:mur, yan}.  Several works have been exclusively devoted to construct soliton, RW and breather solutions of one-component BECs \cite{raj:mur,yan,blud:kono,wen:li,he,loo,Mani} and a few studies have been made to identify the vector soliton and RW solutions of two coupled Gross-Pitaevskii equation (GPE) \cite{raje,raje2, kono,vina,babu,fyu}.  Dynamical evolutions of vector solitons and RWs have been investigated through managing related physical variables, see Refs. \cite{raje,raje2,kono,kanna,vina,fyu}.  Very recently, Babu Mareeswaran et al. \cite{babu} have studied the interaction of the RW with a dark-bright boomeranic soliton of a two-component variant of the NLS equation of relevance to both atomic BECs and nonlinear optics.  Furthermore, they have also examined the robustness of these structures in direct numerical simulation of the original nonautonomous system.  In this paper, we focus our attention to study the characteristics of vector localized matter waves in two and three-component BECs with a cigar-shaped trap.  In this work, we first contruct several possible vector localized solutions for two-component BECs, which can be described by a set of two coupled GPEs with time-dependent scattering length (which is the nonlinearity parameter) and external trap potential.  To capture localized solutions of this model we map the time-dependent two coupled GPEs onto the coupled NLSEs through the similarity transformation method with an integrability condition between the time-dependent scattering length and the external trap potential. In particular, we consider three different forms of traps, namely (i) time-independent expulsive trap, (ii) time-dependent monotonic trap, and (iii) time-dependent periodic trap.  We identify the possibility of vector localized solutions, namely RWs, dark soliton-RW, bright soliton-RW and RW-breather-like structures depending upon the specific values of a particular parameter in the obtained solutions.  We then investigate in detail how the nature of these localized density profiles gets modifed by adjusting the trap parameter.  We depict and analyze the trajectories of the nonautonomous vector RWs.  We also construct the dark-dark RW solution for repulsive-repulsive interaction of the two coupled GPEs and investigate their dynamical evolution when we tune the strength of the trap parameter. In the time-independent trap case our results show that for low values of the trap parameter the vector localized structures are more stable, due to their prolonged existence with respect to time.  By increasing the trap parameter, the localized structures become more and more localized in time, that is localized structures are short-lived, whereas they get stretched or more delocalized in space in a constant density background. In the time-dependent monotonic trap case with high value of the trap parameter, we could not observe any localized structures while $t<0$ because of the form of the nature of potential.  Finally, by replacing the monotonic trap with the periodic trap potential, the localized structures exist on a periodic background when the trap parameter is tuned.  One can also observe that the localized waves lose their stability by way of getting delocalized in space as the potential strength is increased slowly.  While no rigorous quantitative criteria can be identified to locate the point of instability, this can be checked numerically in the original nonautonomous system.

We then construct composite RW solutions for three-component BECs which are described by a coupled set of three GPEs with variable scattering length and external trap potential.  Again, we transform the time-dependent three coupled GPEs to the three coupled NLS equations under the integrability condition through similarity transformation method.  In Ref. \cite{zhao}, the authors restrict the parameter values in the obtained generalized RW solutions and discuss the feasibility of single, two and three composite RWs for three-component NLS systems.  They in fact report that the RW solution exhibits a peculiar structure, namely a four-petaled structure, in contrast to the eye-shaped structure of RW. Motivated by this work, we construct the RW (single, two and three composite RW) solutions and analyze how to control these localized density profiles in three-component BECs.  Furthermore we investigate the characteristics of these localized density profiles when we tune the strength of the trap parameter in the above trap potentials. Our results show that in the case of time-independent trap potential the RW structures maintain their stability when the trap parameter is small and the RW structures become more and more localized in time and stretched (delocalized) in space when we increase the trap parameter. Next we consider the case of time-dependent monotonic trap, and here also we note that the RW structures are more localized in time, delocalized in space and reach the higher density background.  Finally, in the time-dependent periodic trap case, we observe that the RW structures exist on a periodic background when we adjust the trap parameter.  

The paper is organized as follows.  In Sec. II, we present the mean-field model for the two-component BECs, map the quasi-one-dimensional two coupled GPEs to the coupled NLSEs by using the similarity transformation technique which is subjected to a constraint on the forms of the time-dependent scattering length and the external trap potential and obtain the general form of vector RW solutions. The obtained RW solutions can contribute to control the vector localized structures in two-component BECs.  In Sec. III, we identify  different vector localized structures, namely RWs, dark soliton-RW, bright soliton-RW and RW-breather-like waves for different parameteric values and examine their dynamical evolutions in a constant background when we tune the strength of the trap parameter in the time-independent and time-dependent monotonic traps as well as the time-dependent periodic trap.  In Sec. IV, we investigate the trajectories of the nonautonomous RWs.  In Sec. V, we construct the dark-dark RW solution for repulsive-repulsive interaction of two-component BECs and analyze the associated characteristics for the above three different kinds of traps.  In Sec. VI, we consider the mean-field model for the three-component BECs and map the quasi-one-dimensional time-dependent three coupled GPEs to a system of three coupled NLSEs again by using the similarity transformation method. We construct the RW solutions of the three coupled GPEs by considering the different kinds of traps. The obtained RW solutions contribute to control and to understand localized density profiles in three-component BECs.  We illustrate how the localized density profiles such as the single, two and three compoite RWs deform in a constant background when we adjust the trap parameter.  Finally, in Sec. VII, we present a summary of the results and conclusions.

\section{Model and reduction}
At sufficiently low temperatures, the properties of a BEC that is prepared in two hyperfine states of condensate atoms can be described by a set of coupled GP equations of the following form \cite{peth:smi, pitae}, 
\begin{align}
\label{aa1}
i\hbar \frac{\partial \psi_1}{\partial t}=\left[-\frac{\hbar^2}{2m_1} \nabla^2 + V_1(\mathbf r)+U_{11}\vert \psi_1\vert ^2+U_{12}\vert \psi_2\vert ^2\right]\psi_1,\nonumber \\
i\hbar \frac{\partial \psi_2}{\partial t}=\left[-\frac{\hbar^2}{2m_2} \nabla^2 + V_2(\mathbf r)+U_{21}\vert \psi_1\vert ^2+U_{22}\vert \psi_2\vert ^2\right]\psi_2,
\end{align}
where the condensate wave functions $\psi_k(\vec{r},t)$, $k=1,2$, are normalized by the number of atoms for the two components as $N_k=\int \vert \psi_k(\vec{r},t)\vert ^2 d^3r$, where $m_i, i=1,2$, are the masses of the atoms of each components, and $V_i(\mathbf r)$ are the external potentials. The constants $U_{11}$, $U_{22}$ and $U_{12}=U_{21}$ are related to the intraspecies scattering lengths $g_{11}$ and $g_{22}$ and the interspecies scattering length $g_{12}=g_{21}$, respectively, by $U_{ij}=2\pi\hbar^2g_{ij}/m_{ij}\;(i,j=1,2)$, where $m_{ij}=m_im_j/(m_i+m_j)$ is the reduced mass for an atom $i$ and an atom $j$.  The self-interaction is attractive in the case $g_{jj} < 0$ and repulsive for $g_{jj} > 0$,  whereas the interspecies interaction is repulsive when $g_{12} = g_{21} > 0$ and attractive when $g_{12} = g_{21} < 0$.  When both intraspecies interaction and interspecies interaction are all equal and time-dependent \cite{raje}, we can write $g_{11}=g_{22}=g_{12}=g_{21}=a_s(t)$.  The potentials which we consider here are cigar-shaped trap potentials with the elongated axis in the $x$-direction which assume the forms \cite{burger}
\begin{equation}
V_i(\mathbf r, t)  = \frac{m_i}{2} \left [ \omega_i^2 x^2  + \omega^2_{i \perp}(y^2+z^2) \right], \; i=1,2.
\label{eq:3d-trap} 
\end{equation}
When the transverse motions of the condensates are frozen to the ground state in the transverse harmonic trap potential, that is $\omega_{i\perp} \gg \omega_i$, then the system becomes quasi-one-dimensional in nature.  Now, we consider the case $m_1=m_2$ and $\omega_1=\omega_2=\omega$.  Then integrating out the transverse coordinates, Eq. (\ref{aa1}) can be rewritten as a system of quasi one-dimensional coupled GPEs of the form  
\begin{align}
\label{gpe2}
i\frac{\partial \psi_1}{\partial t}+\frac{1}{2}\frac{\partial^2 \psi_1}{\partial x^2}+R(t)(|\psi_1|^2+|\psi_2|^2)\psi_1+\frac{1}{2}\beta^2(t)x^2 \psi_1=0, \nonumber \\
i\frac{\partial \psi_2}{\partial t}+\frac{1}{2}\frac{\partial^2 \psi_2}{\partial x^2}+R(t)(|\psi_1|^2+|\psi_2|^2)\psi_2+\frac{1}{2}\beta^2(t)x^2 \psi_2=0,
\end{align}
where $t$ and $x$ are the temporal and spatial coordinates measured in units $\omega_{\perp}^{-1}$ and $a_{\perp}=\sqrt{\hbar/(m\omega_{\perp})}$, respectively.  Here $R(t)=\frac{2 a_s(t)}{a_0}$, where $a_s(t)$ is the $s$-wave scattering length and $a_0$ is the Bohr radius and $\beta^2(t)=\frac{\omega^2(t)}{\omega_{\perp}^2}$, where $\omega$ is the trap frequency in the axial direction and $\omega_{\perp}$ is the radial trap frequency.  For the expulsive potential the parameter $\beta^2(t)$ is positive ($\beta^2(t) > 0$), and for the confining potential it is negative ($\beta^2(t) < 0$).  The nature of intra- and inter-species interactions is determined by the $s$-wave scattering length, which can be tuned by means of magnetic and optical fields in the vicinity of a Feshbach resonance (FR) \cite{vog:tsa,cout:fre}. This FR technique was utilized experimentally to demonstrate the occurence of bright solitons as the scattering length is tuned from positive to negative values \cite{khay,strec}.  Such possibilities require the scattering length to be a function of time $t$ \cite{moer,abdul}.  On the other hand, the trap frequency in the elongated axis $\omega$ has also been chosen as a function of time $t$ in order to study the characteristics of BECs in the trap.  Therefore, the coefficient of nonlinearity ($R$) and the potential parameter ($\beta$) can be time-dependent.  Eq. (\ref{gpe2}) can be used to describe the management of BECs by suitably choosing the above two time-dependent parameters.  The external potential can be of any form relevant to the experiment.  Few widely considered potentials are (i) linear potential \cite{yang}, (ii) harmonic oscillator potential \cite{belmon}, (iii) optical lattice (OL) potential \cite{ander} and (iv) elliptic function potential \cite{bronski}.  Notably all these potentials can be modulated by a time-dependent function.  These potentials are highly efficient tools for controlling and manipulation of localized matter waves realized in BECs by tuning the external magnetic field and the optically controlled interactions using the FR technique \cite{corn:rob}.

In order to study the dynamics of localized density profiles in two-component BECs, we consider the following similarity transformation \cite{Mani,raje} to map the time-dependent system of two coupled GPEs $(\ref{gpe2})$ to the system of coupled NLSEs,
\begin{equation}
\psi_j(x,t)=r(t)U_j(X,T)\exp[i \theta(x,t)],
\label{a2}
\end{equation}
where $r(t)$ is the amplitude, $T(t)$ is the effective dimensionless time, $X(x,t)$ is the similarity variable and $\theta(x,t)$ is the phase factor which are all to be determined.  To determine these unknown functions we substitute $(\ref{a2})$ into $(\ref{gpe2})$ and obtain a set of polynomial differential equations (PDEs) for the unknown functions.  Now solving these PDEs, we obtain the following relations:
\begin{subequations}
\begin{align}
\label{a3a} r(t)=& \, r_0\sqrt{R(t)}, \\
\label{a3b} \theta(x,t)=& \, -\frac{R(t)_t}{2R(t)}x^2 + br_0^2 R(t)x-\frac{1}{2}b^2r_0^4\int{R^2(t)}dt, \\
\label{a3c} X(x,t)=& \, r_0 R(t)x-br_0^3\int{R^2(t)}dt, \\
\label{a3d} T(t)=& \, \frac{1}{2}r_0^2\int{R^2(t)}dt,
\end{align}
\end{subequations}
where $b$ and $r_0$ are arbitrary constants.  Eq. (\ref{gpe2}) is integrable when the nonlinearity parameter $R(t)$ and the trap potential parameter $\beta(t)$ satisfy the following integrability condition, namely
\begin{equation}
\label{a5}
\frac{d}{dt}\left(\frac{R_t}{R}\right)-\left(\frac{R_t}{R}\right)^2+\beta^2(t)=0,
\end{equation}
which is a Riccati-type equation with dependent variable $(R_t/R)$ and independent variable $t$.  We also note here that even in the presence of a linear Rabi coupling, the integrability condition (\ref{a5}) remains unchanged as the transformation leading to the system (\ref{gpe2}) can be effected through a unitary transformation as was done in Ref. \cite{kanna}. Here also the integrability condition (\ref{a5}) remains unsatisfied for $R$ constant and $\beta^2$ a negative constant, see Ref. \cite{waja}.  Using the above relations $(\ref{a3a})$-$(\ref{a3d})$, the functions $U_j(X,T)$, $j=1,2$, can be found to satisfy the system of coupled NLSEs (Manakov system \cite{mana}) of the form
\begin{equation}
\label{a3}
i \frac{\partial U_j}{\partial T}+\frac{\partial ^2 U_j}{\partial X^2}+2 U_j \sum_{k=1}^{2} |U_k|^2=0, \;\; j=1,2.
\end{equation}
It admits the following special form of solution as shown in Ref. \cite{fabio},
\begin{eqnarray}
\label{soln}
U_1(X,T)=\e^{2i\omega T}\left[\left(\frac{L}{B}\right)a_1+\left(\frac{M}{B}\right)a_2\right], \nonumber \\
U_2(X,T)=\e^{2i\omega T}\left[\left(\frac{L}{B}\right)a_2-\left(\frac{M}{B}\right)a_1\right],
\end{eqnarray}
where $L=\frac{3}{2}-8\omega^2 T^2-2a^2 X^2+8i\omega T+|f|^2 \e^{2aX}$, $M=4f(a X-2 i\omega T-\frac{1}{2})\e^{a X+i \omega T}$ and $B=\frac{1}{2}+8\omega^2 T^2+2a^2 X^2+|f|^2 \e^{2aX}$.  In the above, $a=\sqrt{a_1^2+a_2^2}$, $\omega=a^2$, $a_1$ and $a_2$ are arbitrary real parameters and $f$ is a complex arbitrary constant.  The solution (\ref{soln}) is a semi-rational vector localized solution \cite{fabio}.  A main feature of this solution is that it has both exponential and rational dependence on coordinates.  This general form also yields vector RWs which interact with the soliton waves for particular choice of parameter values.  We note here that the amplitudes of the RWs of each of the components differ from each other unlike the previously reported case of RWs which have same amplitudes in both the components \cite{kono}.  Another interesting property of this solution is that the RWs coexist with dark-bright solitons when we vary the complex arbitrary parameter $f$.  We will discuss these features more elaborately in Section III. 

Regardless of the form of $R(t)$, as long as the condition (\ref{a5}) is satisfied, we obtain the general form of solutions of (\ref{gpe2}) as
\begin{eqnarray}
\label{a7}
\psi_j(x,t) = r_0\sqrt{R(t)}[U_j(X,T)]  \hspace{4cm}  \\
\times \exp \left[{i\left(-\frac{R_t}{R}x^2+br_0^2 R x-\frac{1}{2}b^2r_0^4\int{R^2(t)}dt\right)}\right], j=1,2, \notag
\end{eqnarray}
where $U_j(X,T)$, $j=1,2$, are as given in (\ref{soln}).  The solution (\ref{a7}) accomodates the possibility of generating several localized structures related to RWs, which may be experimentally realizable.  Moreover, the solution helps us to analyze the RW phenomenon relevant to practical situations such as optics and plasmas \cite{kono, mek}.   
     
\section{Characteristics of vector RWs in BECs}
In this section, we shall investigate the characteristics of the semirational, multiparametric vector solutions (\ref{a7}) of the system of time-dependent one-dimensional two coupled GPEs (\ref{gpe2}) with different sets of variable scattering lengths and trap potentials. 
\subsection{Time-independent trap} 
\begin{figure}[!ht]
\begin{center}
\includegraphics[width=0.7\linewidth]{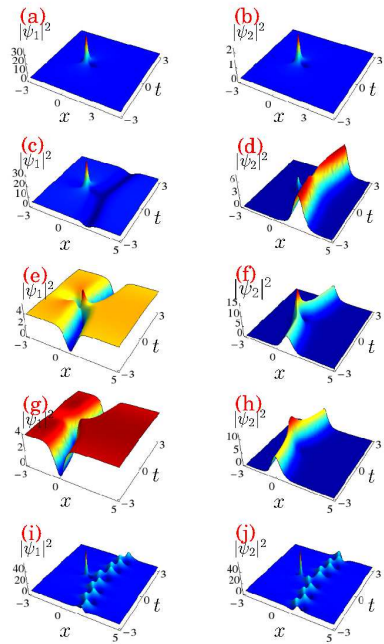}
\end{center}
\caption{(Color online) The density profiles for the two-component BECs $(\ref{gpe2})$ when $R(t)=\sech{(\beta_0 t+\delta)}$ and $\beta^2(t)=\beta_0^2$ for low strength, $\beta_0=0.01$. (a)-(b) Vector RWs for $f=0$, $a_1=2.0$ and $a_2=0.1$. Dark-soliton with RW for $a_1=2.0$ and $a_2=0$ with (c) $f=0.25$, (e) $f=2.5$, (g) $f=15$. Bright-soliton with RW for $a_1=2.0$ and $a_2=0$ with (d) $f=0.25$, (f) $f=2.5$, (h) $f=15$. Panels (i)-(j): Breather-like wave with dark and bright contributions for $f=0.1i$, $a_1=2.5$ and $a_2=2.5$. The other parameters are $r_0=1.0$, $\beta_0=0.1$, $b=0.01$, and $\delta=0.01$.}
\label{fig1}
\end{figure}
\begin{figure}[!ht]
\begin{center}
\includegraphics[width=0.9\linewidth]{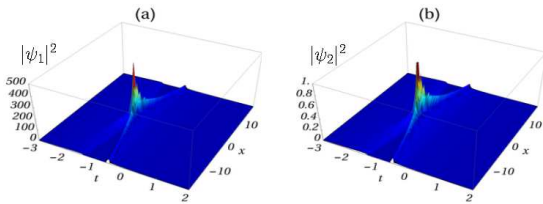}
\end{center}
\caption{(Color online) The density profiles of the RWs in BECs obtained by numerically solving Eq. (\ref{gpe2}) through split-step Crank-Nicolson method for the time-dependent nonlinearity coefficient $R(t)=\sech{(\beta_0 t+\delta)}$ and time-independent trap frequency $\beta(t)^2=\beta_0^2$. The initial condition chosen corresponds to the analytic solution of Figs.~\ref{fig1}(a)-(b).}
\label{num-fig-2}
\end{figure}
\begin{figure}[!ht]
\begin{center}
\includegraphics[width=0.9\linewidth]{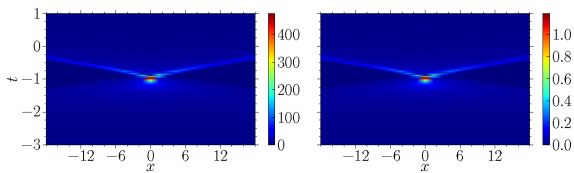}
\end{center}
\caption{(Color online) Contour plots corresponding to Fig.~\ref{num-fig-2}.}
\label{num-fig-2a}
\end{figure}
\begin{figure}[!ht]
\begin{center}
\includegraphics[width=0.7\linewidth]{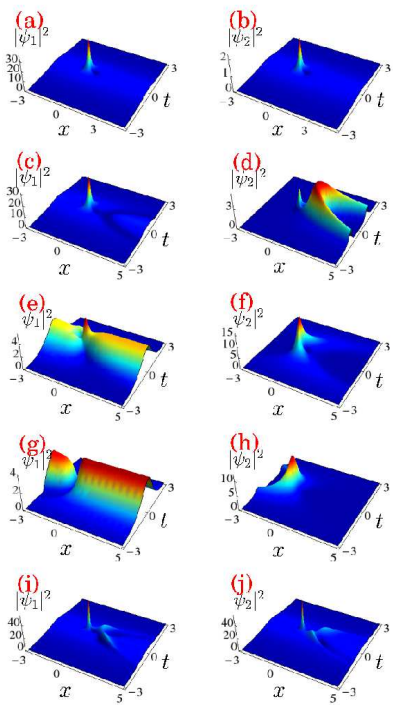}
\end{center}
\caption{(Color online) The density profiles for higher strength trap parameter $\beta_0=1.5$. The other parameters are same as in Fig.~\ref{fig1}.}
\label{fig3}
\end{figure}
To begin with, we consider the case where the trap frequency is a time-independent one, that is $\beta^2(t)=\beta_0^2$. When we substitute this in the integrability condition $(\ref{a5})$ we can obtain the nonlinearity coefficient to be of the form $R(t)=\sech{(\beta_0 t+\delta)}$.  Plugging the chosen form of $\beta(t)$ and $R(t)$ in (\ref{a7}), we find the following RW solution of the time-dependent two coupled GPEs (\ref{gpe2}), that is
\begin{align}
\label{a15}
\psi_j(x,t) = r_0\sqrt{\sech{(\beta_0 t+\delta)}} \, U_j(X,T) \, \exp\left(i \chi(x,t)\right),  
\end{align}
where $j = 1, 2, $ and
\begin{eqnarray}
\chi(x,t) = \left[b r_0^2 \sech{(\beta_0 t+\delta)}x+\frac{(\beta_0^2 x^2-b^2r_0^4)\tanh{(\beta_0 t+\delta)}}{2\beta_0}\right], \nonumber
\end{eqnarray}
and $U_1(X,T)$, and $U_2(X,T)$ are as given in Eq. (\ref{soln}) with $X(x,t)=r_0  \sech{(\beta_0 t+\delta)}x-b r_0^3/\beta_0 \tanh{(\beta_0 t+\delta)}$, and $T(t)=r_0^2/(2\beta_0) \tanh{(\beta_0 t+\delta)}$.  Fig.~\ref{fig1} shows various types of density profiles of the nonlinear localized matter waves for the time-independent trap frequency $\beta^2(t)=\beta_0^2$ and time-dependent scattering length $R(t)=\sech{(\beta_0 t+\delta)}$.  The particular case, namely $f=0$ in Eq. (\ref{a15}) for $U_j(X,T)$ as specified by Eq. (\ref{soln}) is the RW solution, that is the Peregrine soliton.  The solution representing the vector RWs is depicted in terms of $|\psi_1(x,t)|^2$ and $|\psi_2(x,t)|^2$ when $\delta=0.01$, $a_1=2.0$ and $a_2=0.1$ in Figs. \ref{fig1}(a)-(b).  The components of RWs reach the maximum amplitudes approximately at $t\approx0$.  The amplitudes are different in each of the components and are found to be $|\psi_1(x,t)|^2\approx30$ and $|\psi_2(x,t)|^2\approx3$.  Further, the RW is unstable: after reaching the maximum value, and it disappears approximately at $t\approx1$. 

Next we consider the case $f\neq0$, $a_1=2.0$ and $a_2=0$ in Eq. (\ref{a15}) and analyze the existence of the RW and its interaction with a soliton which propagates with a nonconstant speed, for different values of $|f|$.  In the asymptotic limit  $T\to\pm\infty$, the ratios $L(X,T)/B(X,T)$ and $M(X,T)/B(X,T)$ in Eq. (\ref{soln}) describe the dark and bright contributions, respectively, and the individual contributions of the dark shape $L/B$ and bright shape $M/B$ appear separately when $a_2=0$. 

When $f=0.25$, Figs. \ref{fig1}(c)-(d) show the interaction of RW with the dark soliton and bright soliton in $|\psi_1(x,t)|^2$ and $|\psi_2(x,t)|^2$ components, respectively. When we increase the value of $f$ in the parametric RW solution (\ref{a15}), the single RW merges with the dark-bright soliton as depicted in Figs. \ref{fig1}(e)-(h).  Here we notice that the density of the first component $|\psi_1(x,t)|^2$ decreases while the density of the second component $|\psi_2(x,t)|^2$ increases.  The RW completely merges with dark- and bright-soliton when $f=2.5$ as presented in Figs. \ref{fig1}(e)-(f).  At $f=15$,  the RW cannot be identified while the resulting dark-bright distribution appears as a boomeron-type soliton as shown in Figs. \ref{fig1}(g)-(h).  From this discussion we conclude that when the value of $|f|$ is low, the RW and dark-bright solitons separate from each other and when the value of $|f|$ is increased the RW and dark-bright solitons merge.  These facts can be observed in each of the components. 

In the above, we have considered only real values for the complex parameter $f$.  Now we consider the case when the parameter $f$ is complex.  In this case the solution (\ref{a15}) reveals that the matter waves behave like a breather.  Figs. \ref{fig1}(i)-(j) show breather-like waves resulting from the interference between the dark and bright contributions for $f=0.1i$, $a_1=2.5$ and $a_2=2.5$.  When we decrease the value of $|f|$, the RW and breather-like wave get separated from each other which is not shown here.  

For the evidence of presence of the RWs further, we have also performed a direct numerical simulation of (\ref{gpe2}) with the aid of the split-step Crank-Nicolson method using an initial wave function which is the same as the function (\ref{a15}) and with space step $dx=0.015$ and time step $dt=0.001$~\cite{Muruganandam+}.  The computer generated density profile of the RWs and the corresponding contour plots are presented in Figs.~\ref{num-fig-2} and \ref{num-fig-2a} with the parameters chosen as $r_0=1.0$, $c_1=0.01$, $\beta_0=0.1$, $f=0$, $a_1=2.0$, $a_2=0.1$ and $\delta=0.01$ which are the same as that of Figs.~\ref{fig1}(a)-(b). The analytically obtained results are in good agreement with the numerically computed for the emergence of the RWs.  We have also verified numerically the existence of RW with dark and bright soliton of (\ref{gpe2}) as well, replicating Figs.~\ref{fig1}.

In the above investigation we have fixed the strength of the trap parameter to be rather low at $\beta_0=0.1$. The interaction of RWs with bright and dark solitons represents the exchange of condensate atoms between the RW and the soliton/breather and this exchange keeps the structural stability against the attractive interatomic interaction and trap potential.  Now we tune the strength of the paratemeter $\beta_0$ to $1.5$.  Here also we can see the RWs in each component.  The density fluctuations with constant density background become more and more localized in time as shown in Figs. \ref{fig3}(a)-(b).  From Figs. \ref{fig3}(c)-(d) we observe that the amplitude of the RW remains constant whereas the dark and bright solitons exhibit more and more localization in time and their amplitudes decrease in space.  The figures reveal that the exchange of atoms between RW and soliton stretches in space which tells us that the atoms that constitute the RW and solitons become more and more localized in time and delocalized in space.  The delocalization of condensate atoms in the density background is position dependent which can be observed in Figs. \ref{fig3}(e)-(h).  The RW with breather-like wave in each of the components is more and more localized in time as seen in Figs. \ref{fig3}(i)-(j).  Thus, comparing Figs. \ref{fig1} and \ref{fig3}, one can conclude that when the interaction strength $\beta_0$ is increased, the density profile of the condensate atoms suffers a collapse in space to a constant density background whereas it is more localized in time.

\subsection{Time-dependent monotonic trap}
Now, we consider a time-dependent trap and investigate how it affects the vector localized structures.  The time-dependent trap frequency which we consider is of the form $\beta^2(t)=\left({\beta_0^2}/{2} \right)\left[1-\tanh\left({\beta_0 t/}{2}\right)\right]$. The integrability condition (\ref{a5}), gives the time-dependent interaction term to be $R(t)=1+\tanh \left({\beta_0 t}/{2}\right)$.  The general expression of the parametric solution (\ref{a7}) becomes  
\begin{align}
\label{a22}
\psi_j(x,t)&=&r_0\sqrt{1+\tanh\Big(\frac{\beta_0}{2}t\Big)}  \, U_j(X,T) \, \exp\left(i \chi(x,t)\right),    
\end{align}%
where $j = 1, 2,$ and
\begin{eqnarray}
\chi(x,t) & = & \left[\frac{\beta_0\sech^2\left(\frac{\beta_0 t}{2}\right)x^2}{4\left[1+\tanh(\frac{\beta_0 t}{2})\right]}-br_0^2\left[1+\tanh{\left(\frac{\beta_0 t}{2}\right)}\right]x \right. \nonumber \\ & & 
\left. +\frac{b^2r_0^4\left(\beta_0 t+2\log\left[\cosh(\frac{\beta_0 t}{2})-\tanh(\frac{\beta_0 t}{2})\right]\right)}{\beta_0}\right], \nonumber
\end{eqnarray}
and $U_1(X,T)$, and $U_2(X,T)$ are as given in equation (\ref{soln}) with $X(x,t)=r_0\left(1+\tanh \left({\beta_0 t}/{2}\right)\right)x-2 b r_0^3/\beta_0 \left(\beta_0 t+2\log\left[\cosh(\frac{\beta_0 t}{2})-\tanh(\frac{\beta_0 t}{2})\right]\right)$, and $T(t)=r_0^2/\beta_0 \left(\beta_0 t+2\log\left[\cosh(\frac{\beta_0 t}{2})-\tanh(\frac{\beta_0 t}{2})\right]\right)$.
\begin{figure}[!ht]
\begin{center}
\includegraphics[width=0.7\linewidth]{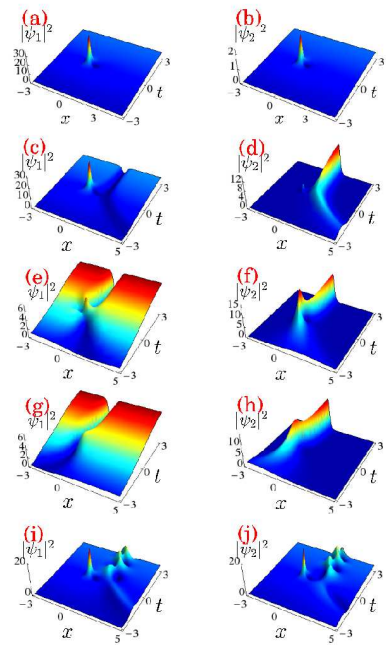}
\end{center}
\caption{(Color online) As in Fig. \ref{fig1} with $\beta_0=0.7$ for $R(t)=1+\tanh \left({\beta_0 t}/{2}\right)$ and $\beta(t)^2=\left({\beta_0^2}/{2} \right)\left[1-\tanh\left({\beta_0 t/}{2}\right)\right]$.}
\label{fig5}
\end{figure}
\begin{figure}[!ht]
\begin{center}
\includegraphics[width=0.6\linewidth]{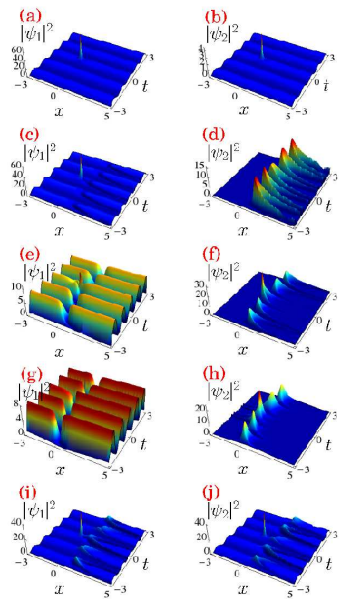}
\end{center}
\caption{(Color online) As in Fig. \ref{fig1} with $\beta_0=2.5$ for $R(t)=1+\cos{(2 \beta_0 t)}$ and $\beta(t)^2=2\beta_0^2[1+3\tan^2(\beta_0 t)]$.}
\label{fig7}
\end{figure}

The qualitative nature of the various localized matter waves for $R(t)=1+\tanh \left({\beta_0 t}/{2}\right)$ and $\beta^2(t)=\left({\beta_0^2}/{2} \right)\left[1-\tanh\left({\beta_0 t/}{2}\right)\right]$ turns out to be the same as in the previous case (Fig. \ref{fig1}) when the strength of the trap parameter $\beta_0=0.1$ and so we do not display the outcome here separately.  Figs. \ref{fig5} show various density profiles of the matter nonlinear localized waves exhibited by the above forms of $\beta^2(t)$ and $R(t)$ for $\beta_0=0.7$, while the other parameters are as in Figs. \ref{fig1}.  In Figs. \ref{fig5}(a)-(b), we observe that the vector RWs exist on a high density background when $\beta_0=0.7$.  From Figs. \ref{fig5}(c)-(j), we can observe that for $t\leq 0$, the localized matter waves tend to bend and disappear because of decreasing lifetime of the condensate atoms inside the trap.  Figs. \ref{fig5}(c) and Fig. \ref{fig5}(d) show the RW interaction with the bending profile of dark-soliton of $|\psi_1(x,t)|^2$ and the bending profile of bright-soliton of $|\psi_2(x,t)|^2$, respectively, for $\beta_0=0.7$.   Figs. \ref{fig5}(e)-(f) represent the bending vector dark-bright soliton together with a single RW for each of the components.  The bending profile of boomeronic type soliton is shown in Figs. \ref{fig5}(g)-(h).  Figs. \ref{fig5}(i)-(j) show the modified structure of breather-like waves resulting from the interference of the dark and bright contributions with the RW for $\beta_0=0.7$.  In this case also we notice that the density fluctuations of the condensate atoms are more localized in time and delocalized in space, while for $t<0$ the RW as well as the soliton/breather gets collapsed in each of the components. 
\subsection{Time-dependent periodic trap}
Finally, we consider the temporal periodic modulation of the trap potential in the form $\beta^2(t)=2\beta_0^2[1+3\tan^2(\beta_0 t)]$ for the two-component BECs.  The time-dependent interatomic interaction term is $R(t)=1+\cos{(2 \beta_0 t)}$ which is consistent with the integrability condition (\ref{a5}).  The BEC system exhibits the matter wave solution of the form
\begin{align}
\label{a24}
\psi_j(x,t)=r_0\sqrt{1+\cos{(2\beta_0 t)}}  \, U_j(X,T) \, \exp\left(i \chi(x,t)\right), 
\end{align}
where $j = 1, 2,$ and
\begin{eqnarray}
\chi(x,t)& = & \left[\beta_0 \tan{(\beta_0 t)}x^2+2 b r_0^2  \cos{(\beta_0 t)^2} x \right. \notag \\ & &
\left.-\frac{b^2 r_0^4(12\beta_0 t+8\sin{(2\beta_0 t)}+\sin{(4\beta_0 t)})}{16\beta_0}\right],\notag
\end{eqnarray}
where $U_1(X,T)$, and $U_2(X,T)$ are given by Eq. (\ref{soln}) with $X(x,t)=2r_0\cos{(\beta_0 t)}^2 x-br_0^3/8\beta_0(12\beta_0 t+8\sin{(2\beta_0 t)}+\sin{(4\beta_0 t)})$, and $T(t)=r_0^2/(16\beta_0)(12\beta_0 t+8\sin{(2\beta_0 t)}+\sin{(4\beta_0 t)})$.
   
Here also we obtained the various localized structures when we vary the parameter $f$, by fixing $\beta_0=0.1$.  The vector localized structures for the time-dependent periodic trap $\beta^2(t)=2\beta_0^2[1+3\tan^2(\beta_0 t)]$ and temporal periodic modulation of the scattering length $R(t)=1+\cos{(2 \beta_0 t)}$ are similar to the previous two cases and so we do not display the outcome here separately.  Next we consider a high value of $\beta_0$. The parameters are kept the same as in Figs. \ref{fig1} with $\beta_0=2.5$. Figs. \ref{fig7}(a)-(b) show the vector RWs exhibited on a periodic wave background for $f=0$.  As $f$ is increased from 0.25, we notice that the amplitude of the RW dissolves into the approaching dark soliton on a periodic background in $|\psi_1(x,t)|^2$ as shown in Figs. \ref{fig7}(c), (e) and (g). For similar parametric tuning, RWs appear on the bright solitonic background that folds to localize in positive space in $|\psi_2(x,t)|^2$ as shown in Figs. \ref{fig7}(d) and (f). A further increase in $f$, the amplitude of RW diminish, with bright soliton localizing in the positive space as presented in Fig. \ref{fig7}(h). Also the bright and dark contributions, for complex $f$, exhibit density fluctuations of the RW alongside breather like structures as shown in Figs. \ref{fig7}(i)-(j).
 
In summary we conclude the following features which are noted in the above three cases of trap potentials when the trap parameter $\beta_0$ is varied.  (i) Time-independent trap: The vector localized structures exhibit more localization in time and delocalization in space.  (ii) Time-dependent monotonic trap: The vector localized structures (RW and soliton/breather profiles) interact only when $t\geq0$ and for $t<0$ the localized structures get collapsed (disappeared) due to the nature of attractive potential.  (iii) Time-dependent periodic trap: In this case also the characteristic interaction of the RW and soliton/breather profiles repeat as in the previous two cases, but on a periodic background.  
\section{Trajectories of the nonautonomous RW}
In this section, we study the characteristics of RW, namely the evolution of its hump, width (distance between the two valleys) and the nature of the trajectory analytically.  The trajectory of RW can be described by the motion of the hump and the valley \cite{ling}.  When $f=0$ in (\ref{a7}) one can obtain the vector RWs of Eq. (\ref{gpe2}).  With the exact solution of the vector RW solution, we can find the position of the peak ($x_h$) and the two valleys ($x_{v_1}$ and $x_{v_2}$) which appear in the atomic density profiles.  For this purpose, we apply the extremum theorem to equation (\ref{a7}), namely $[\partial \vert \psi_j(x,t)\vert^2/\partial x]_{x=x_c}=0$, and obtain
\begin{align}
\label{cube}
(br_0^2\tilde{R}-x_c R(t))\big[(3+4(a_1^2+a_2^2)r_0^4(3a_1^2+3a_2^2-b^2r_0^2)\tilde{R}^2  \nonumber \\  +8(a_1^2+a_2^2)br_0^4x_c\tilde{R}R(t)-4(a_1^2+a_2^2)r_0^2x_c^2R^2(t))\big]=0, 
\end{align}
where $\tilde{R}=\int R^2(t) dt $.  Consequently, we have either one of the two following possibilities:
\begin{subequations}
\begin{eqnarray}
br_0^2\tilde{R}-x_c R(t)=0, \\
(3+4(a_1^2+a_2^2)r_0^4(3a_1^2+3a_2^2-b^2r_0^2)\tilde{R}^2 \\ +8(a_1^2+a_2^2)br_0^4x_c\tilde{R}R(t)-4(a_1^2+a_2^2)r_0^2x_c^2R^2(t))&=&0. \nonumber
\end{eqnarray}
\end{subequations}
By solving the above expressions ($x_c$ denotes $x_h$, $x_{v_1}$ or $x_{v_2}$), we can find the positions of the hump and the two valleys with respect to time.  The expression for the position of the hump ($x_h$) is given by
\begin{equation}
\label{xhump}
x_h = \frac{b r_0^2 \tilde{R}}{R(t)},
\end{equation}
and the expressions for the positions of the two valleys ($x_{v_1}$, $x_{v_2}$) are given by
\begin{figure}[!ht]
\begin{center}
\includegraphics[width=0.99\linewidth]{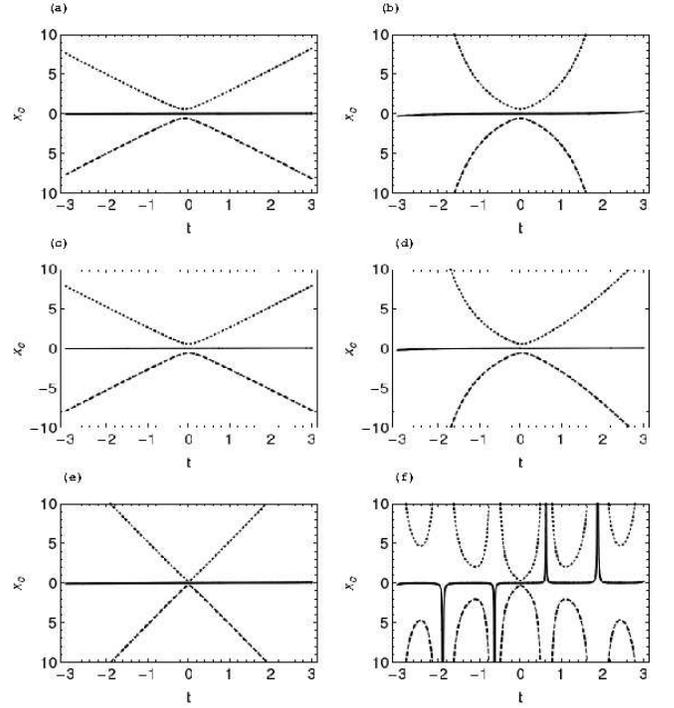}
\end{center}
\caption{(Color online) The trajectory of the RW (corresponding to the $|\psi_1|^2$ component) when (a) $\beta_0=0.1$, (b) $\beta_0=1.5$ for $R(t)=\sech{(\beta_0 t+\delta)}$, (c) $\beta_0=0.1$, (d) $\beta_0=1.5$ for $R(t)=1+\tanh \left({\beta_0 t}/{2}\right)$ and (e) $\beta_0=0.1$, (f) $\beta_0=2.5$ for $R(t)=1+\cos{(2\beta_0 t)}$.  The vertical axes $x_c$ denotes $x_h$, $x_{v_1}$ and $x_{v_2}$.  The solid line represents the hump of RW $x_h$, dashed and dotted lines represent the motions of the two valleys $x_{v_1}$ and $x_{v_2}$. The other parameters are the same as in Fig. \ref{fig1}(a).} 
\label{fig8}
\end{figure} 
\begin{eqnarray}
\label{xval}
 x_{v_1,v_2}&=&\frac{2 (a_1^2 + a_2^2) b r_0^4  \tilde{R} R(t)}{(2 (a_1^2 + a_2^2) r_0^2 R^2(t))} \\&& \pm \frac{\sqrt{3} \sqrt{(a_1^2 + a_2^2) r_0^2 (1 + 4 (a_1^2 +a_2^2)^2 r_0^4  \tilde{R}^2) R^2(t)}}{(2 (a_1^2 + a_2^2) r_0^2 R^2(t))}\nonumber.
\end{eqnarray}
\begin{table*}[]
\centering
\caption{The expressions for the hump and the two valleys of RW atomic density profiles for different forms of $R(t)$.  Here $\alpha_1=\frac{4(a_1^2+a_2^2)^2r_0^4\tanh^2{(\beta_0 t+\delta)}}{\beta_0^2}$, $\alpha_2=\beta_0 t+2\log\left[\cosh{(\beta_0 t/2)}-\tanh{(\beta_0 t/2)}\right]$, $\alpha_3=1+\tanh{(\beta_0 t/2)}$ and $\alpha_4=12\beta_0 t+8\sin{(2\beta_0 t)}+\sin{(4t\beta_0)}$.}
\label{table}
\begin{tabular}{|c|c|c|}\hline
  R(t) & $x_h$ & $x_{v_1,v_2}$   \\ \hline
  $\sech{(\beta_0 t+\delta)}$  & $\frac{br_0^2\sinh{(\beta_0 t+\delta)}}{\beta_0}$ &  $\frac{br_0^2\sinh{(\beta_0 t+\delta)}}{\beta_0}\pm\frac{\sqrt{3}\cosh^2{(\beta_0 t+\delta)}\sqrt{a_1^2+a_2^2r_0^2\sech^2{(\beta_0 t+\delta)}\left(1+\alpha_1\right)}}{2(a_1^2+a_2^2)r_0}$  \\ \hline
  $1+\tanh \left({\beta_0 t}/{2}\right)$  & $\frac{2br_0^2\alpha_2}{\beta_0\alpha_3}$ & $\frac{1}{2r_0^2\alpha_3^2}\left(\frac{4br_0^4\alpha_2\alpha_3}{\beta_0} \pm \frac{\sqrt{3(a_1^2+a_2^2)r_0^2\left(1+\frac{16(a_1^2+a_2^2)r_0^4\alpha_2^2}{\beta_0^2}\right)\alpha_3^2}}{a_1^2+a_2^2}\right)$   \\ \hline
  $1+\cos{(2\beta_0 t)}$  & $\frac{br_0^2\sec^2{(\beta_0 t)}\alpha_4}{16\beta_0}$ & $\frac{\sec^2{(\beta_0 t)}}{16r_0^2}\left(\frac{br_0^4\alpha_4}{\beta_0}\pm \frac{4\sec^2{(\beta_0 t)}\sqrt{3(a_1^2+a_2^2)r_0^2\cos^4{(\beta_0 t)}\left(1+\frac{(a_1^2+a_2^2)r_0^4\alpha_4^2}{16\beta_0^2}\right)}}{a_1^2+a_2^2}\right)$  \\ \hline
\end{tabular}
\end{table*}
The forms of the hump and valleys of the RW for different $R(t)$ are tabulated (see Table \ref{table}) and plotted in Fig. {\ref{fig8}}.  
The atomic density at the maximum of the RW is given by
\begin{eqnarray}
|\psi_1|^2_{max}=\frac{a_1^2r_0^2(9+4(a_1^2+a_2^2)^2r_0^4\tilde{R}^2)R(t)}{1+4(a_1^2+a_2^2)r_0^4\tilde{R}^2},
\end{eqnarray}
and at the minimum of RW, it is given by
\begin{eqnarray}
|\psi_1|^2_{min}=\frac{4a_1^2(a_1^2+a_2^2)^2r_0^6\tilde{R}^2R(t)}{1+4(a_1^2+a_2^2)r_0^4\tilde{R}^2}.
\end{eqnarray}
Similarly $|\psi_2|^2_{max,min}$ can be calculated.  The width of the RW (distance between two valleys) evolves with time as
\begin{eqnarray}
W(t)=\frac{\sqrt{3(a_1^2+a_2^2)r_0^2(1+4(a_1^2+a_2^2)^2r_0^4\tilde{R}^2)R^2(t)}}{(a_1^2+a_2^2)r_0^2R^2(t)}.
\end{eqnarray}
The above expressions provide relevant information on the evolution of the hump and valleys of the atomic density profiles.  Figs. \ref{fig8}(a)-(b) show the trajectory of RW (corresponding to the $|\psi_1|^2$ component) for $R(t)=\sech{(\beta_0 t+\delta)}$.  The trajectory of the hump of the RW $(x_h)$ is shown by a solid line while the motion of the two valleys are represented by dotted and dashed lines. Fig. \ref{fig8}(a) gives the trajectories of RW for $\beta_0=0.1$.  From the figure, we can observe that $|\psi_1|^2_{max}$ travels almost in a straight line in the neighborhood of the origin where the valleys travel in an $'X'$ shaped path.  Near $t=0$, the two valleys come closer to the hump position and give rise to the localized RW structure.  Now increasing the trap parameter to $\beta_0=1.5$, in Fig. \ref{fig8}(b), we can observe that the positions of the two valleys are well separated at $t=-2$ and they approach closer to each other very rapidly than in the previous case and give rise to the locallized RW structure near $t=0$.  Soon after, we can find that the trajectories of both the valleys rapidly deviate from each other. In a similar way, we have plotted the trajectories for two different parameter values of RW for $R(t)=1+\tanh \left({\beta_0 t}/{2}\right)$ as shown in Figs. \ref{fig8}(c)-(d).  Similar to the previous case, an increase in $\beta_0$ causes rapid changes in the positions of the two valleys.  Also we can find that the trajectories of the two valleys are not symmetric with respect to time for higher values of $\beta_0$ (Fig. \ref{fig8}(d)).  Similarly the trajectories of the RW for the time periodic trap case are shown in Figs. \ref{fig8}(e)-(f) when $R(t)=1+\cos{(2\beta_0 t)}$.  From this figure, we can find that for a lower value of $\beta_0$, the trajectories of the extrema of the RW behave similar to the previous case.  However, for increased value of $\beta_0$, we find that $|\psi_1|^2_{max}$ itself follows a periodic trajectory.  From the trajectories of the valleys, we can find that the two valleys come closer to each other periodically and at $t=0$, the two valleys are found to be the closest.  Thus there arises a periodic wave background around the RW around $t=0$.  
\section{Nonautonomous dark-dark RWs in BECs}
In the previous sections, we have considered the model for attractive-attractive interatomic interaction of two-component BECs and analyzed how the vector localized density profiles behave, namely the vector RW and its interaction with dark-bright solitons and breather-like waves have been analyzed.  However, BECs exhibit few other further interesting localized structures as well.  The other predominant structure is the dark-dark RW.  Here we construct the dark-dark RW solutions of the two-component BECs.  To study this we  consider the model for repulsive-repulsive interatomic interaction which is modelled by the following time-dependent two coupled quasi one-dimensional GPEs \cite{peth:smi},
\begin{align}
\label{dgpe2}
i\frac{\partial \psi_1}{\partial t}+\frac{1}{2}\frac{\partial^2 \psi_1}{\partial x^2}-R(t)(|\psi_1|^2+|\psi_2|^2)\psi_1+\frac{1}{2}\beta^2(t)x^2 \psi_1=0, \nonumber \\
i\frac{\partial \psi_2}{\partial t}+\frac{1}{2}\frac{\partial^2 \psi_2}{\partial x^2}-R(t)(|\psi_1|^2+|\psi_2|^2)\psi_2+\frac{1}{2}\beta^2(t)x^2 \psi_2=0,
\end{align}
where $\psi_j$, $j=1,2$, are the condensate wave functions for the two components and the other variables and parameters are as defined under Eq. (\ref{gpe2}).  The vector GPE (\ref{dgpe2}) can also be transformed to a different system of coupled NLSEs through the same similarity transformation as given in (\ref{a2}).  We find the unknown functions $r(t), \theta(x,t)$, and $X(x,t)$ have exactly the same form as already given in (\ref{a3a})-(\ref{a3c}), but $T(t)= -\frac{1}{2}r_0^2\int{R^2(t)}dt$.  Eq. (\ref{dgpe2}) is integrable when the nonlinearity parameter $R(t)$ and the trap potential parameter $\beta(t)$ satisfy the same integrability condition (\ref{a5}).  The function $U_j(X,T)$ can be found to satisfy the coupled NLSEs with defocusing of the form
\begin{equation}
\label{da3}
i \frac{\partial U_j}{\partial T}-\frac{\partial ^2 U_j}{\partial X^2}+ 2 \left(\sum_{k=1}^{2} |U_k|^2\right) U_j=0, \;\; j=1,2. 
\end{equation}
Eq. (\ref{da3}) admits the following form of special solutions as shown in \cite{jhli}
\begin{eqnarray}
\label{darksoln}
U_1(X,T)&=& \rho \exp{(4 i\rho^2 T)}\nonumber \\ && \left(1+\frac{4 g^2(-1+i(-s X-s^2 T + g^2 T))}{(g^2+s^2)(g^2(X+s T)^2+g^4T^2+1)}\right), \nonumber \\
U_2(X,T)&=& \rho \exp{(i s X+i s^2 T+4 i \rho^2 T)} \\ && \left(1+\frac{4 g^2(-1+i(-sX-s^2 T+g^2 T))}{(g^2+s^2)(g^2(X+s T)^2+g^4T^2+1)}\right),\nonumber
\end{eqnarray}
where $g=\pm \sqrt{-4\rho^2-s^2+2\sqrt{(2 \rho^2)^2+4s^2\rho^2}}$ and the criterion for the existence of RW is $s^2<8\rho^2$.  When we vary the parameter $s$ in the above expression, we identify two different interesting localized structures, namely the dark-dark RWs, whose amplitudes drop to zero, and the four-petal configuration dark-dark RWs.  In the following we will discuss in detail how the nature of the dark-dark RW structures get deformed in a constant density background when we change the parameter $\beta_0$.

Regardless of the form of $R(t)$, as long as the condition (\ref{a5}) is satisfied, we obtain the dark-dark RW solutions of $(\ref{dgpe2})$ in the form
\begin{eqnarray}
\label{da7}
\psi_j(x,t)&=&r_0\sqrt{R(t)}[U_j(X,T)]  \\&& 
\hspace{-1.3cm}\exp \left[{i\left(-\frac{R_t}{R}x^2+br_0^2 Rx-\frac{1}{2}b^2r_0^4\int{R^2(t)}dt\right)}\right], j=1,2, \nonumber
\end{eqnarray}
where $U_j(X,T)$, $j=1,2$ is the solution of coupled NLSEs (\ref{da3}) which is given in (\ref{darksoln}).

\subsection{Characteristics of dark-dark RWs in BECs}
We begin our studies with time-independent trap $\beta^2(t)=\beta_0^2$ and $R(t)=\sech{(\beta_0 t+\delta)}$ which is consistent with the integrability condition (\ref{a5}).  With this form of $R(t)$, we can obtain the dark-dark RW solutions of the time-dependent two coupled GPEs from Eq. (\ref{da7}). 
\begin{figure}[!ht]
\begin{center}
\includegraphics[width=0.85\linewidth]{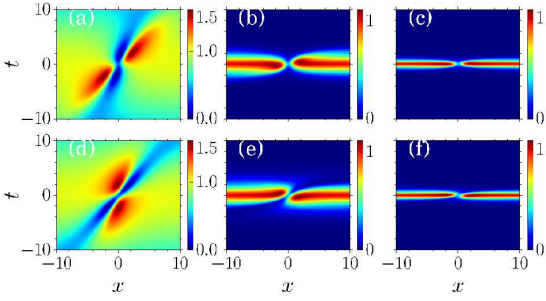}
\end{center}
\caption{(Color online) Contour plots of the density profiles (a)-(c) $|\psi_1(x,t)|^2$ and (d)-(f) $|\psi_2(x,t)|^2$ of vector dark RWs for $R(t)=\sech{(\beta_0 t+\delta)}$ and $\beta^2(t)=\beta_0^2$. The parameter $\beta_0$ is varied as (a), (d) $\beta_0=0.1$, (b), (e) $\beta_0=1.0$, and (c), (f) $\beta_0=2.5$.  The other parameters are $\rho=1$, $s=1.0$, $r_0=1.0$, $b=0.01$, and $\delta=0.01$.}
\label{fig10}
\end{figure}
\begin{figure}[!ht]
\begin{center}
\includegraphics[width=0.85\linewidth]{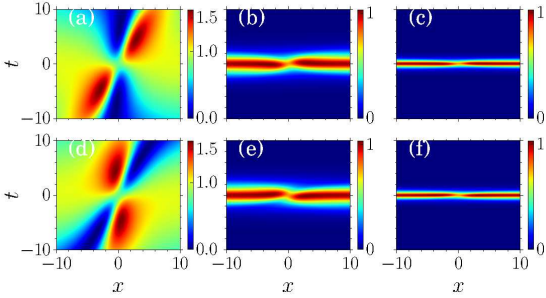}
\end{center}
\caption{(Color online) Contour plots of the density profiles (a)-(c) $|\psi_1(x,t)|^2$ and (d)-(f) $|\psi_2(x,t)|^2$ of vector four-petal configuration dark RWs for $R(t)=\sech{(\beta_0 t+\delta)}$ and $\beta^2(t)=\beta_0^2$. The parameter $\beta_0$ is varied as (a), (d) $\beta_0=0.1$, (b), (e) $\beta_0=1.0$, and (c), (f) $\beta_0=2.5$.  The other parameters are $\rho=1$, $s=0.5$, $r_0=1.0$, $b=0.01$, and $\delta=0.01$.}
\label{fig11}
\end{figure}
Figs.~\ref{fig10}(a) and \ref{fig10}(d) show the corresponding contour plots of the density profiles of dark-dark matter RWs for the strength of the trap parameter $\beta_0=0.1$.  When we tune the interaction strength $\beta_0$ the density profiles of dark RWs (the density fluctuations of condensate atoms in the constant density background) become more and more localized in time as shown in Figs. \ref{fig10}(b) and \ref{fig10}(c) for the $|\psi_1(x,t)|^2$ component and $|\psi_2(x,t)|^2$ component in Figs. \ref{fig10}(e) and \ref{fig10}(f), respectively.  Fig.~\ref{fig11} displays the density profiles of dark-dark matter RWs in a four-petal configuration for the same above forms of $\beta^2(t)$ and $R(t)$ as a function of $\beta_0$.  The qualitative nature of the dark RW pair in the four-petal configuration is shown in Figs. \ref{fig11}(a) and \ref{fig11}(d) for the strength of the trap parameter $\beta_0=0.1$. As $\beta_0$ is increased, we obtain a deformed structure, that is the density fluctuations are more and more localized in time and delocalized in space as seen from Figs. \ref{fig11}(b) and \ref{fig11}(c) for the $|\psi_1(x,t)|^2$ component and Figs. \ref{fig11}(e) and \ref{fig11}(f) for the $|\psi_2(x,t)|^2$ component, respectively.  Thus the nature of the density fluctuations can be controlled by varying the interaction strength between the atoms.

Next, we consider the case of the time-dependent trap frequency $\beta^2(t)=\left({\beta_0^2}/{2} \right)\left[1-\tanh\left({\beta_0 t/}{2}\right)\right]$. The integrability condition (\ref{a5}) fixes the time-dependent interaction term to be $R(t)=1+\tanh \left({\beta_0 t}/{2}\right)$. 
\begin{figure}[!ht]
\begin{center}
\includegraphics[width=0.85\linewidth]{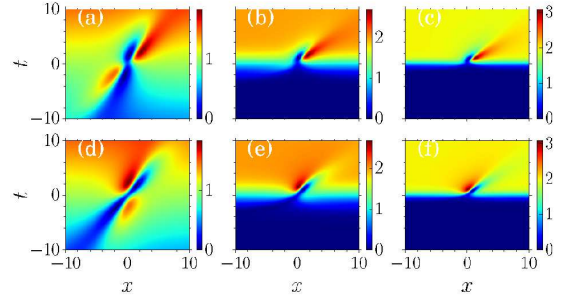}
\end{center}
\caption{(Color online) Contour plots of the density profiles (a)-(c) $|\psi_1(x,t)|^2$ and (d)-(f) $|\psi_2(x,t)|^2$ of vector dark RWs for $R(t)=1+\tanh \left({\beta_0 t}/{2}\right)$ and $\beta^2(t)=\left({\beta_0^2}/{2} \right)\left[1-\tanh\left({\beta_0 t/}{2}\right)\right]$. The parameter $\beta_0$ is varied as (a), (d) $\beta_0=0.1$, (b), (e) $\beta_0=1.0$, and (c), (f) $\beta_0=2.5$.  The other parameters are $\rho=1$, $s=1.0$, $r_0=1.0$, $c_1=0.01$, and $\delta=0.01$.}
\label{fig13}
\end{figure}
In Fig. \ref{fig13}, we present a pair of dark RWs for these choices of $R(t)$ and $\beta^2(t)$.  When $\beta_0=0.1$ the dark-dark RWs are as shown in Figs. \ref{fig13}(a) and \ref{fig13}(d).  By tuning the value of the interaction strength $\beta_0$, the structure of the dark-dark RWs collapses when $t \leq 0$ and the density fluctuations of atoms get settled at different constant density backgrounds as seen from Figs. \ref{fig13}(b) and \ref{fig13}(c) for the $|\psi_1(x,t)|^2$ component and Figs. \ref{fig13}(e) and \ref{fig13}(f) for the $|\psi_2(x,t)|^2$ component, respectively. 
\begin{figure}[!ht]
\begin{center}
\includegraphics[width=0.85\linewidth]{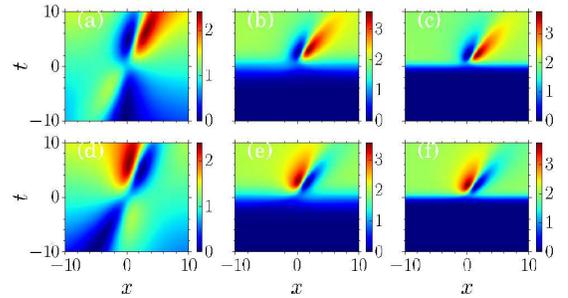}
\end{center}
\caption{(Color online) Contour plots of the density profiles (a)-(c) $|\psi_1(x,t)|^2$ and (d)-(f) $|\psi_2(x,t)|^2$ of vector dark RWs in the deformed structure of four-petal configuration for $R(t)=1+\tanh \left({\beta_0 t}/{2}\right)$ and $\beta^2(t)=\left({\beta_0^2}/{2} \right)\left[1-\tanh\left({\beta_0 t/}{2}\right)\right]$. The parameter $\beta_0$ is varied as (a), (d) $\beta_0=0.1$, (b), (e) $\beta_0=1.0$, and (c), (f) $\beta_0=2.5$.  The other parameters are $\rho=1$,$s=1.0$, $r_0=1.0$, $b=0.01$, and $\delta=0.01$.}
\label{fig15}
\end{figure}
The density profiles of the corresponding vector RWs in a four-petal configuration are presented in Fig. \ref{fig15}.  The vector dark RWs in the deformed structure of four-petal configuration for $\beta_0=0.1$ is shown in Figs. \ref{fig15}(a) and \ref{fig15}(d).  When the strength of trap parameter $\beta_0$ is increased, we observe that the structures are modified which are shown in Figs. \ref{fig15}(b) and \ref{fig15}(e). On further increasing the value of the strength of the trap parameter $\beta_0$ to $2.5$, we notice that a pair of density profiles acquire a modified structure and reach a higher density background when $t\geq 0$ which is displayed in Figs. \ref{fig15}(c) and \ref{fig15}(f).

Finally, we consider the temporal periodic modulation of the trap potential $\beta^2(t)=2\beta_0^2[1+3\tan^2(\beta_0 t)]$ for the two-component BECs.  In accordance with the integrability condition (\ref{a5}) the time-dependent interatomic interaction term is $R(t)=1+\cos{(2 \beta_0 t)}$.
\begin{figure}[!ht]
\begin{center}
\includegraphics[width=0.85\linewidth]{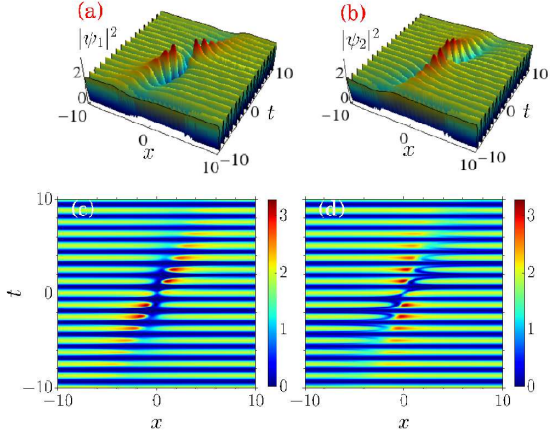}
\end{center}
\caption{(Color online) The density profiles (a) $|\psi_1(x,t)|^2$ and (b) $|\psi_1(x,t)|^2$ of vector dark RWs for $R(t)=1+\cos{(2 \beta_0 t)}$ and $\beta^2(t)=2\beta_0^2[1+3\tan^2(\beta_0 t)]$ and (c) and (d) are their corresponding contour plots. The parameters are $\rho=1$, $s=1.0$, $\beta_0=2.5$, $r_0=1.0$, $b=0.01$, and $\delta=0.01$.}
\label{fig16}
\end{figure}
\begin{figure}[!ht]
\begin{center}
\includegraphics[width=0.85\linewidth]{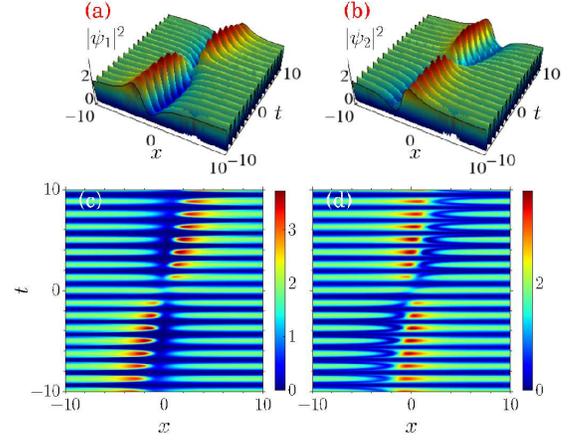}
\end{center}
\caption{(Color online) The density profiles (a) $|\psi_1(x,t)|^2$ and (b) $|\psi_1(x,t)|^2$ of vector dark RWs in a four-petal configuration for $R(t)=1+\cos{(2 \beta_0 t)}$ and $\beta^2(t)=2\beta_0^2[1+3\tan^2(\beta_0 t)]$ and (c) and (d) are their corresponding contour plots. The parameters are $\rho=1$, $s=0.5$, $\beta_0=2.5$, $r_0=1.0$, $b=0.01$, and $\delta=0.01$.}
\label{fig17}
\end{figure}
Figs. (\ref{fig16})-(\ref{fig17}) represent the density profiles of the vector dark RWs that exist on a periodic background for the above forms of $R(t)$ and $\beta^2(t)$. 
\section{RWs in three component Bose-Einstein condensates}
In the mean-field approximation, the dynamics of a three-component BECs in a quasi-one-dimensional approximation with equal time-dependent interaction strengths and in the presence of external time-dependent harmonic potential is described by the following dimensionless non-autonomous three-coupled GPEs \cite{pitae,peth:smi}, namely
\begin{align}
\label{gpe3}
i\frac{\partial \psi_j}{\partial t}+\frac{1}{2}\frac{\partial^2 \psi_j}{\partial x^2}+\sum_{k=1}^{3}R(t)|\psi_k|^2\psi_j+\frac{1}{2}\beta^2(t)x^2 \psi_j=0, 
\end{align}
where $\psi_j$, $j=1,2,3$, are the condensate wave functions for three components, $t$ and $x$ are the temporal and spatial coordinates, respectively.  As in the previous cases, the nonlinearity parameter $R(t)$ describes the variation of scattering length and can be controlled well by Feshbach resonance \cite{vog:tsa,cout:fre,corn:rob} and $\beta^2(t)$ is the trap potential parameter.  The time-dependent three coupled GPEs (\ref{gpe3}) is, in general, non-integrable and in order to study the dynamics of (\ref{gpe3}), we can map the three component GPEs $(\ref{gpe3})$ to a set of three coupled NLSEs under the similarity transformation (\ref{a2}).  The unknown functions of $r(t)$, $\theta(x,t)$, $R(t)$ and $\beta(t)$ are as given in Eqs. (\ref{a3a}), (\ref{a3b}) and (\ref{a5}), respectively  and the function $U_j(X,T)$ can be found to satisfy the set of three coupled NLSEs of the form as
\begin{align}
\label{ta3}
i \frac{\partial U_j}{\partial T}+\frac{\partial ^2 U_j}{\partial X^2}+2 U_j \sum_{k=1}^{3} |U_k|^2=0, \;\; j=1,2,3,
\end{align}
where $X$ and $T$ are as defined in Eqs. (\ref{a3c}) and (\ref{a3d}).  Eq. (\ref{ta3}) admits the following form of solution \cite{zhao}  
\begin{eqnarray}
\label{3soln}
U_1(X,T) &=& 1-\frac{H_1(X,T)}{G_1(X,T)}\exp\left[{i\frac{9T}{2}-i\frac{X}{\sqrt{2}}}\right], \nonumber \\
U_2(X,T) &=& 1-\frac{H_2(X,T)}{G_2(X,T)}\frac{\exp{[5iT]}}{\sqrt{2}}, \\
U_3(X,T) &=& 1-\frac{H_3(X,T)}{G_3(X,T)}\exp\left[{i\frac{9T}{2}+i\frac{X}{\sqrt{2}}}\right], \nonumber 
\end{eqnarray}
where $H_j(X,T)$ and $G_j(X,T)$, j=1,2,3, are sixth order polynomial functions in $X$ and $T$ as given in the Appendix of Ref. \cite{zhao} by Zhao and Liu.  Again regardless of what $R(t)$ is, as long as the condition (\ref{a5}) is satisfied, we obtain the solution of $(\ref{gpe3})$ as 
\begin{eqnarray}
\label{ta7}
\psi_j(x,t)&=&r_0\sqrt{R(t)}[U_j(X,T)]  \\&& 
\hspace{-1.3cm}\exp \left[{i\left(-\frac{R_t}{R}x^2+br_0^2 Rx-\frac{1}{2}b^2r_0^4\int{R^2(t)}dt\right)}\right], j=1,2,3 \nonumber
\end{eqnarray}
where $U_j(X,T)$, $j=1,2,3$ is the above solution (\ref{3soln}) of the coupled NLSEs (\ref{ta3}). The complete integrability of Eq. (\ref{ta3}) was studied in \cite{zhao}. As we are interested in the RW solutions of (\ref{gpe3}) we invoke only this solution for (\ref{ta3}) as reported in \cite{zhao}.  We do not reproduce the explicit forms of $H_j$ and $D_j$, $j=1,2,3$, in $(\ref{3soln})$ here due to their lengthy nature but use these expressions in our further analysis.         
\subsection{Characteristics of the nonautonomous RW in three-component BECs}
We discuss the characteristics of RW solutions of time-dependent one-dimensional system of three coupled GPEs with  different kinds of variable scattering lengths and trap potentials, as in the case of two-component BECs in the earlier sections. 
\subsubsection{Time-independent trap} 
Substituting the nonlinearity parameter $R(t)=\sech{(\beta_0 t+\delta)}$ and time-independent trap parameter $\beta^2(t)=\beta_0^2$ in $(\ref{ta7})$, we can obtain the RW solution of the time-dependent three coupled GPEs, $\psi_j(x,t) = r_0\sqrt{\sech{(\beta_0 t+\delta)}} \, U_j(X,T) \, \exp(i\chi(x,t)), \;j = 1, 2, 3$, with $\chi(x,t)$ as given in (\ref{a15}) and $U_1(X,T)$, $U_2(X,T)$ and $U_3(X,T)$ are as given in (\ref{3soln}) with $X(x,t)=r_0  \sech{(\beta_0 t+\delta)}x-b r_0^3/\beta_0 \tanh{(\beta_0 t+\delta)}$, and $T(t)=r_0^2/(2\beta_0) \tanh{(\beta_0 t+\delta)}$.  This solution with four free parameters namely $A_1, A_2, A_3$ and $A_4$ is very lengthy and so we do not give the explicit expression here and analyze the results only graphically.  With suitable restrictions on the parameters $A_1$, $A_2$, $A_3$ and $A_4$ we obtain different RW structures in this system (\ref{gpe3}).  For example (i) when $A_3=0$, $A_4=0$ and $A_1$, $A_2 \neq 0$, we obtain a single RW in each component, (ii) $A_4=0$, and $A_1$, $A_2$, $A_3 \neq 0$, we extract two composite RWs in each component and (iii) when $A_4 \neq 0$, we obtain three composite RWs in the $x-t$ plane.  In the following, we discuss these three cases one by one by choosing the trap parameter as $\beta_0=0.1$.  

\begin{figure}[!ht]
\begin{center}
\includegraphics[width=0.99\linewidth]{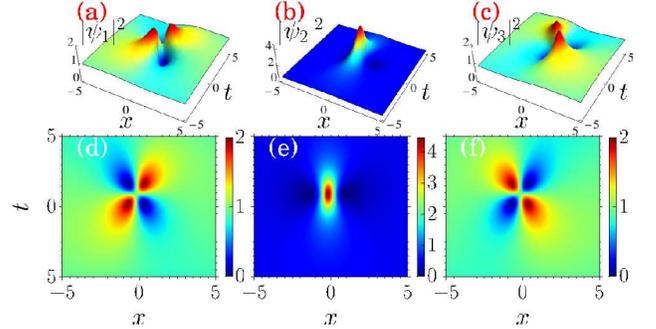}
\end{center}
\caption{(Color online) The density profiles of the RWs of three-component BEC system (\ref{gpe3}). (a), (c) Four-petaled RW in the components $|\psi_1(x,t)|^2$ and $|\psi_3(x,t)|^2$ and (b) single RW structure in $|\psi_2(x,t)|^2$ for $R(t)=\sech{(\beta_0 t+\delta)}$ and $\beta^2(t)=\beta_0^2$. (d), (e) and (f) are the corresponding contour plots of (a), (b) and (c). The parameters are chosen as $A_1=1$, $A_2=5$, $A_3=0$, $A_4=0$, $r_0=1.0$, $\beta_0=0.1$, $b=0.01$, and $\delta=0.01$.}
\label{fig18}
\end{figure}

\begin{figure}[!ht]
\begin{center}
\includegraphics[width=0.99\linewidth]{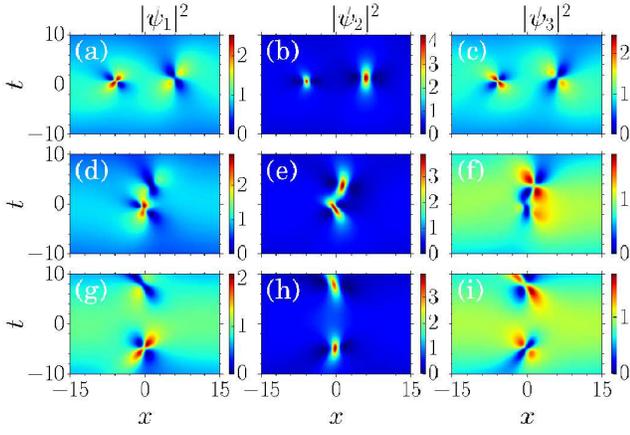} 
\end{center}
\caption{(Color online) Contour plots corresponding to two composite RWs: The parameters (a)-(c) $A_1=-2$, $A_2=30$, $A_3=6$ and $A_4=0$, (d)-(f) $A_1=80$, $A_2=30$, $A_3=6$ and $A_4=0$, and (g)-(i) $A_1=120$, $A_2=30$, $A_3=6$ and $A_4=0$.  The other parameters are the same as in Fig. \ref{fig18}.}
\label{fig19}
\end{figure}

\begin{figure}[!ht]
\begin{center}
\includegraphics[width=0.99\linewidth]{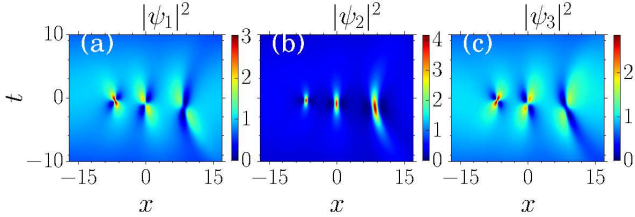}
\end{center}
\caption{(Color online) Contour plots corresponding to three composite RWs (a)-(c) for the parameters $A_1=10$, $A_2=50$, $A_3=2$ and $A_4=5$. The other parameters are same as in Fig. \ref{fig18}.}
\label{fig20}
\end{figure}

\begin{figure}[!ht]
\begin{center}
\includegraphics[width=0.99\linewidth]{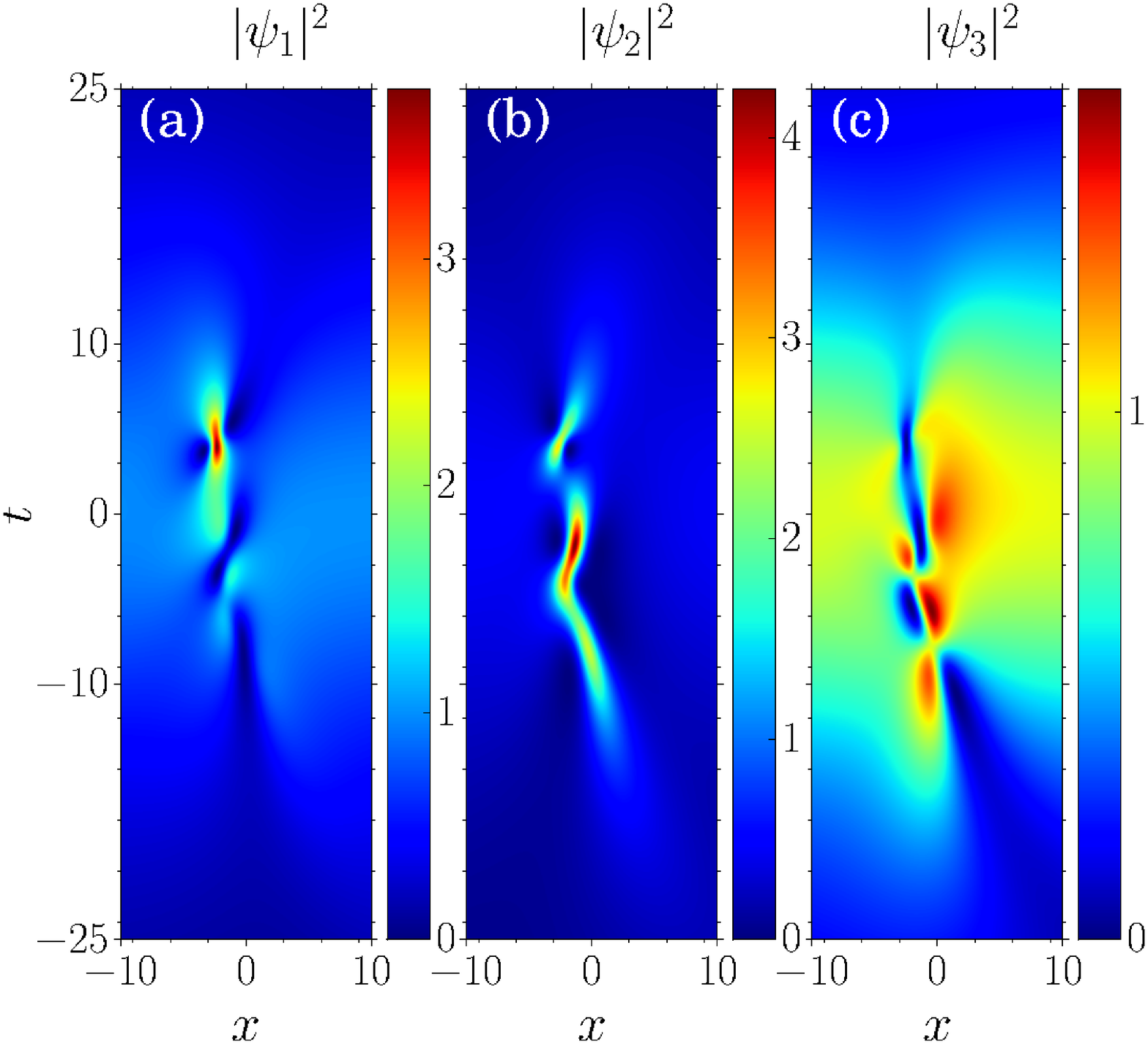} 
\end{center}
\caption{(Color online) Contour plots corresponding to the interactions of three composite RWs (a)-(c) for the parameters $A_1=150$, $A_2=250$, $A_3=120$ and $A_4=50$. The other parameters are same as in Fig. \ref{fig18}.}
\label{fig21}
\end{figure}

Figs. \ref{fig18}-\ref{fig21} show three different kinds of matter RWs for time-independent harmonic trap potential parameter $\beta^2(t)=\beta_0^2$ and time-dependent scattering length parameter $R(t)=\sech{(\beta_0 t+\delta)}$. Here we consider the parameters $A_3=0$ and $A_4=0$ in the obtained solution.  Fig. \ref{fig18} represents the density profiles of single RWs, $|\psi_1(x,t)|^2$, $|\psi_2(x,t)|^2$ and $|\psi_3(x,t)|^2$ of (\ref{gpe3}).  We can see that the density distribution shapes of the matter RWs in $|\psi_1(x,t)|^2$ and $|\psi_3(x,t)|^2$ are quite distinct from $|\psi_2(x,t)|^2$.  From Figs. \ref{fig18}(a) and \ref{fig18}(c), we observe two humps and two valleys around a center and the position of the center is almost equal to that of the background, giving the appearance of a four-petaled structure.  This structure can be clearly seen in the contour plots, Figs. \ref{fig18}(d) and \ref{fig18}(f), respectively.  On the other hand, the density distribution in the $|\psi_2|^2$ component appears similar to that of the RW in a single component system for which there are one hump and two valleys, as presented in Figs. \ref{fig18}(b) and \ref{fig18}(e).  In each component, the RW disappears when it reaches the maximum, thus revealing the unstable nature of RWs. We also note that the amplitudes in their components $|\psi_1(x,t)|^2$, $|\psi_3(x,t)|^2 \approx 2$ and $|\psi_2(x,t)|^2 \approx 4$.  In Fig. \ref{fig18}, the second panel shows the corresponding contour plots of (a), (b) and (c).

Next we restrict only the parameter $A_4=0$ in the obtained RW solutions for (\ref{gpe3}).  Here we observe two composite RWs in each component.  When $A_1 \leq 0$, two composite RWs exist at a certain time, as shown in Figs. \ref{fig19}(a)-(c).  The four-petaled RWs arise in $|\psi_{1,3}|^2$ which are shown in Figs. \ref{fig19}(a) and (c) and the single RW emerges in the $|\psi_2|^2$ component which is presented in Fig. \ref{fig19}(b).   Two composite RWs arise at two different times, when $A_1 \geq 0$, which is shown in Figs. \ref{fig19}(d)-(i).   The interactions of the two matter RWs are also shown in these figures, when $A_1=80$. When we increase the parameter to $A_1=120$, the two RWs separate from each other which is demonstrated in Figs. \ref{fig19}(h)-(i). Thus each of the components $\psi_j(x,t)$, $j=1,2,3$, show two composite matter RWs in the case of the three-component GPEs (\ref{gpe3}).  

When all the parameters $A_1, A_2, A_3$ and $A_4$ are nonzero, we obtain three composite RWs in each of the components.  At $A_4=-1$, we can observe three composite RWs at a particular time, as shown in Figs. \ref{fig20}(a)-(c).  The four-petaled structure of RWs appear in $|\psi_{1,3}|^2$ components as displayed in Figs. \ref{fig20}(a) and \ref{fig20}(c) and three separated RWs arise in the $|\psi_2|^2$ component as exhibited in Fig. \ref{fig20}(b).  Next, we choose $A_4$ to be positive and demonstrate the interactions of the three composite RWs as shown in Figs. \ref{fig21}(a)-(c). Thus each one of the components $\psi_j(x,t)$, $j=1,2,3$, show triple vector matter RWs.  
\begin{figure}[!ht]
\begin{center}
\includegraphics[width=0.9\linewidth]{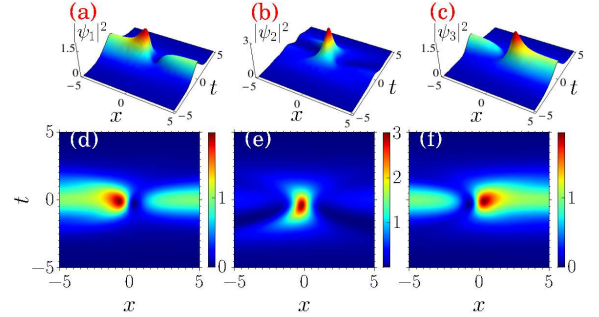}
\end{center}
\caption{(Color online) (a), (c) Four-petaled RW and (b) RW structure, and (d), (e) and (f) are the corresponding contour plots of (a), (b) and (c). The parameters are same as in Fig. \ref{fig18} with $\beta_0=1.2$.}
\label{fig22}
\end{figure}
\begin{figure}[!ht]
\begin{center}
\includegraphics[width=0.9\linewidth]{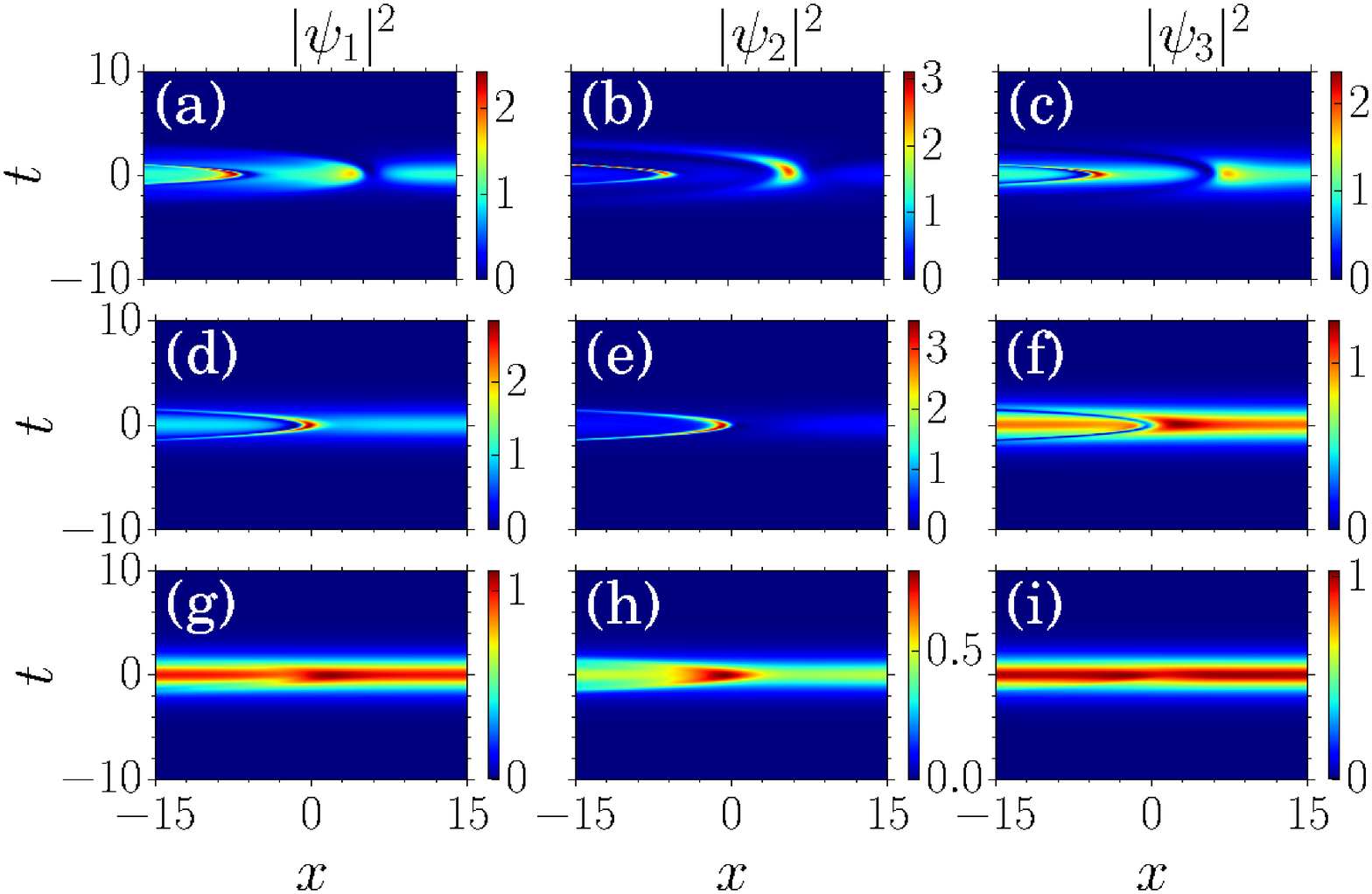} 
\end{center}
\caption{(Color online) Contour plots corresponding to two composite RWs. The parameters are same as in Fig. \ref{fig19} with $\beta_0=1.2$.}
\label{fig23}
\end{figure}
\begin{figure}[!ht]
\begin{center}
\includegraphics[width=0.9\linewidth]{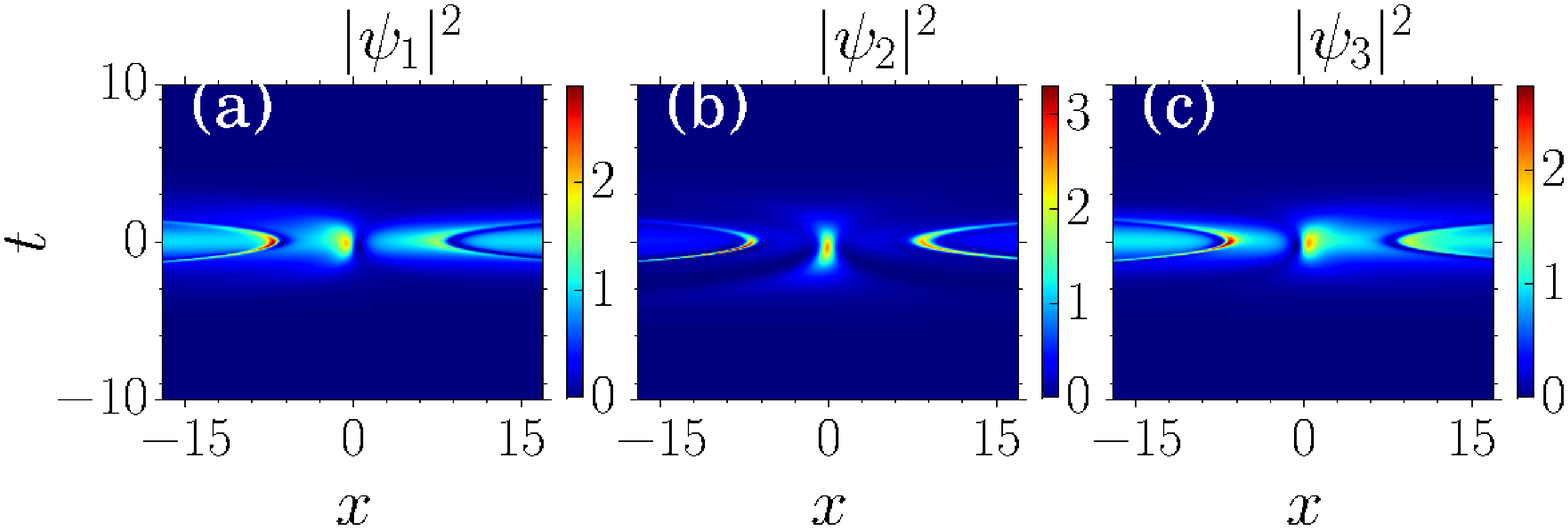}
\end{center}
\caption{(Color online) Contour plots corresponding to three composite RWs. The parameters are same as in Fig. \ref{fig20} with $\beta_0=1.2$.}
\label{fig24}
\end{figure}
\begin{figure}[!ht]
\begin{center}
\includegraphics[width=0.9\linewidth]{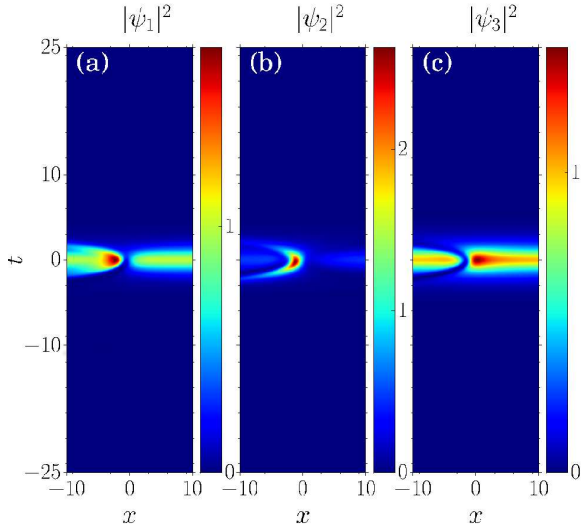} 
\end{center}
\caption{(Color online) Contour plots corresponding to the interactions of three composite RWs. The parameters are same as in Fig. \ref{fig21} with $\beta_0=1.2$.}
\label{fig25}
\end{figure}

Figs. \ref{fig22}-\ref{fig25} depict the qualitative nature of a different set of localized density profiles of (\ref{gpe3}).  Typical distributions of $|\psi_1(x,t)|^2$, $|\psi_2(x,t)|^2$ and $|\psi_3(x,t)|^2$ components show how the nature of RW structures get modified in a constant density background when we adjust the trap parameter $\beta_0$.  For this investigation we fixed all the parameters as the same as in the previous discussion and we only increase the value of the parameter $\beta_0$ to $1.2$.  In Figs. \ref{fig22}(a)-(c), we observe the four-petaled RW structure and the RW compresses in time and stretches (delocalize) in space.  The corresponding contour plots are presented in Figs. \ref{fig22}(d)-(f).  We then examine the two composite RWs when we alter the strength of the trap parameter.  The modified structure of the two composite RWs is as shown in Figs. \ref{fig23}(a)-(i). This figure reveals that the density distribution of condensate atoms get more and more localized in time and delocalized in space.  We now analyze the case of the three composite RWs.  In Figs. \ref{fig24}(a)-(c), we observe that three composite RWs become more localized in time and delocalization of condensate atoms are position dependent.  We note here that the RW structure gets more and more localilzed in time and delocalized in space and two RWs of them disappear which are displayed in Figs. \ref{fig25}(a)-(c).  In the above, we have monitored how the RW structures get modified in a constant density background which corresponds to the phenomenon of condensate atoms getting exchanged between the RW and its background.  This exchange of atoms allows the system to maintain the constant background of the RW even while varying the trap parameter.

\subsubsection{Time-dependent monotonic trap}
Next, we consider the form of time-dependent trap potential $\beta^2(t)=\left({\beta_0^2}/{2} \right)\left[1-\tanh\left({\beta_0 t/}{2}\right)\right]$.  The integrability condition (\ref{a5}) gives the time-dependent interaction term $R(t)=1+\tanh \left({\beta_0 t}/{2}\right)$.  After substituting these forms of $R(t)$ and $\beta^2(t)$ in (\ref{ta7}), the form of RW solution reads  $\psi_j(x,t)$=$r_0\sqrt{1+\tanh\Big(\frac{\beta_0}{2}t\Big)}  \, U_j(X,T) \, \exp(\chi(x,t)), \;j = 1, 2, 3$, with $\chi(x,t)$ as given in (\ref{a22}) and $U_1(X,T)$, $U_2(X,T)$ and $U_3(X,T)$ are given in Eq. (\ref{3soln}) with $X(x,t)=r_0\left(1+\tanh \left({\beta_0 t}/{2}\right)\right)x-2 b r_0^3/\beta_0 \left(\beta_0 t+2\log\left[\cosh(\frac{\beta_0 t}{2})-\tanh(\frac{\beta_0 t}{2})\right]\right)$, and $T(t)=r_0^2/\beta_0 \left(\beta_0 t+2\log\left[\cosh(\frac{\beta_0 t}{2})-\tanh(\frac{\beta_0 t}{2})\right]\right)$.

\begin{figure}[!ht]
\begin{center}
\includegraphics[width=0.9\linewidth]{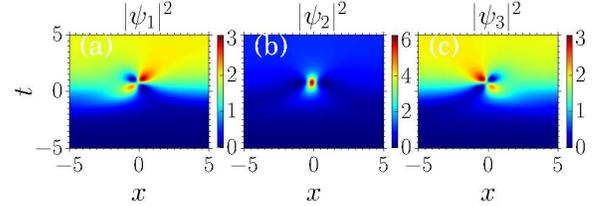}
\end{center}
\caption{(Color online) Contour plots: (a), (c) Four-petaled RW and (b) RW structure for $R(t)=1+\tanh \left({\beta_0 t}/{2}\right)$ and $\beta^2(t)=\left({\beta_0^2}/{2} \right)\left[1-\tanh\left({\beta_0 t/}{2}\right)\right]$. The parameters are $A_1=1$, $A_2=5$, $A_3=0$, $A_4=0$,$r_0=1.0$, $\beta_0=1.2$, $b=0.01$, and $\delta=0.01$.}
\label{fig26}
\end{figure}
\begin{figure}[!ht]
\begin{center}
\includegraphics[width=0.9\linewidth]{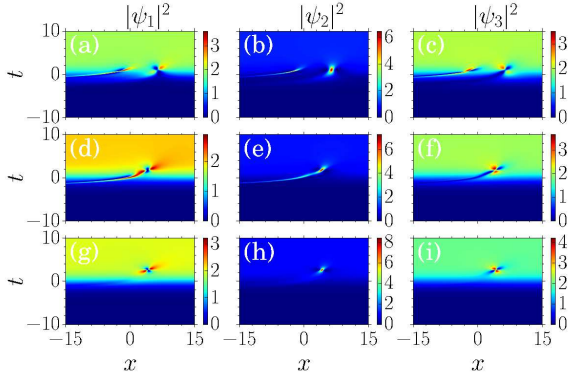} 
\end{center}
\caption{(Color online) Contour plots corresponding to two composite RWs: The parameters (a)-(c) $A_1=-2$, $A_2=30$, $A_3=6$ and $A_4=0$, (d)-(f) $A_1=80$, $A_2=30$, $A_3=6$ and $A_4=0$, and (g)-(i) $A_1=120$, $A_2=30$, $A_3=6$ and $A_4=0$.  The other parameters are same as in Fig. \ref{fig26}.}
\label{fig27}
\end{figure}
\begin{figure}[!ht]
\begin{center}
\includegraphics[width=0.9\linewidth]{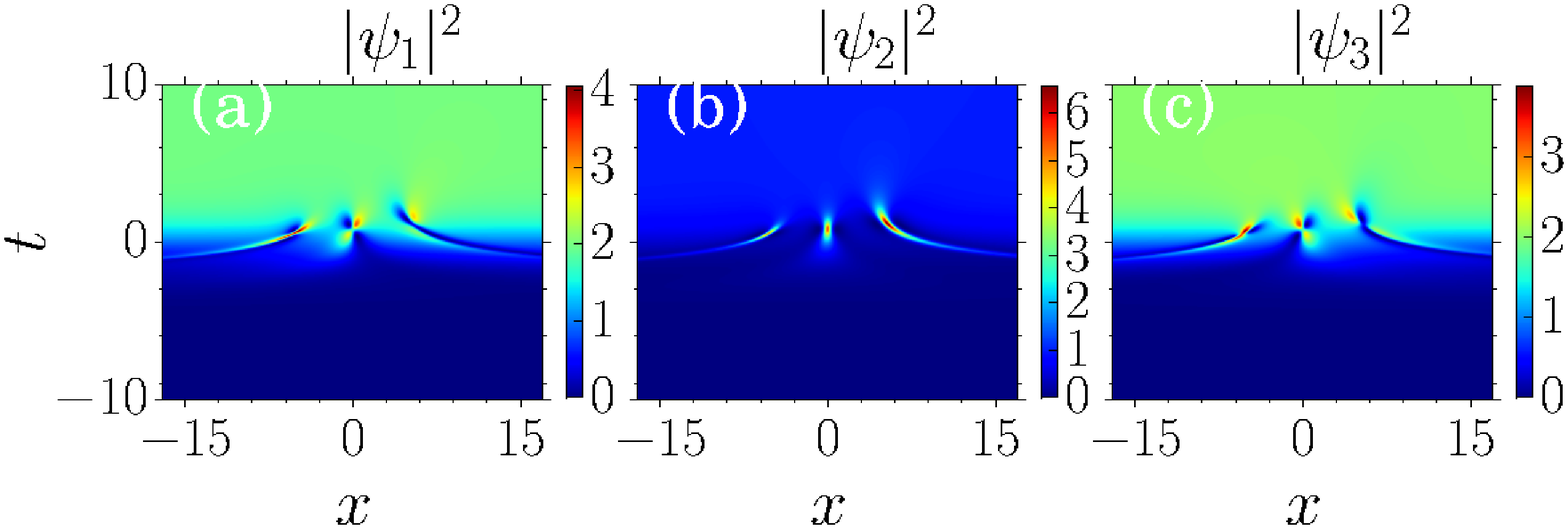} 
\end{center}
\caption{(Color online) Contour plots corresponding to three composite RWs: The parameters (a)-(c) $A_1=10$, $A_2=50$, $A_3=2$ and $A_4=5$. The other parameters are same as in Fig. \ref{fig26}.}
\label{fig28}
\end{figure}
\begin{figure}[!ht]
\begin{center}
\includegraphics[width=0.9\linewidth]{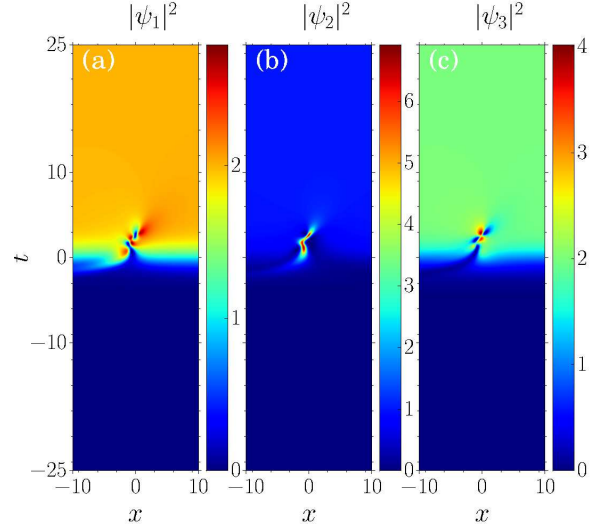} 
\end{center}
\caption{(Color online) Contour plots corresponding to the interactions of three composite RWs: The parameters (a)-(c) $A_1=150$, $A_2=250$, $A_3=120$ and $A_4=50$. The other parameters are same as in Fig. \ref{fig26}.}
\label{fig29}
\end{figure}

In Figs. \ref{fig26}-\ref{fig29} we display the qualitative nature of different localized density profiles for time-dependent harmonic trap potential $\beta(t)^2=\left({\beta_0^2}/{2} \right)\left[1-\tanh\left({\beta_0 t/}{2}\right)\right]$ and interatomic interaction $R(t)=1+\tanh \left({\beta_0 t}/{2}\right)$.  By keeping the trap parameter $\beta_0=0.1$, we obtain the RW density profiles which are the same as what we observed in the previous case and so we do not repeat the discussion here.  We observe the following features when we alter the trap parameter.  Here, we consider all the other parameters as the same as in the previous case and we vary the parameter $\beta_0$ from $0.1$ to $1.2$.  In Figs. \ref{fig26}(a) and \ref{fig26}(c), we note that the four-petaled RW structure delocalize in space and reach a higher density background when $t \geq 0$.  In Fig. \ref{fig26}(b), we observe that the RW get more localized in the constant background. The deformed two composite RWs under change in the strength of the trap parameter are as shown in Figs. \ref{fig27}(a)-(i).  The three composite RWs bend in the constant background as presented in Figs. \ref{fig28}(a)-(c).  The interactions of the three RW structure make them to delocalize in space which is represented in Figs. \ref{fig29}(a)-(c).  We conclude here that when we tune the strength of the trap parameter the density profiles of the RW structures get delocalized in space and reach a high density background.


\subsubsection{Time-dependent periodic trap}
Finally, we consider the temporal periodic modulation of the trap potential $\beta^2(t)=2\beta_0^2[1+3\tan^2(\beta_0 t)]$ for the three-component BECs.  In accordance with the integrability condition (\ref{a5}), we have the time-dependent interatomic interaction term as $R(t)=1+\cos{(2 \beta_0 t)}$.  Using the above forms into (\ref{ta7}), we have the RW solution of the form $\psi_j(x,t)=r_0\sqrt{1+\cos{(2\beta_0 t)}}  \, U_j(X,T) \, \exp(\chi(x,t)), \;j = 1, 2, 3,$ with $\chi(x,t)$ as given in (\ref{a24}) and $U_1(X,T)$, $U_2(X,T)$ and $U_3(X,T)$ as given in (\ref{3soln}) with $X(x,t)=2r_0\cos^2{(\beta_0 t)} x-br_0^3/8\beta_0(12\beta_0 t+8\sin{(2\beta_0 t)}+\sin{(4\beta_0 t)})$, and $T(t)=r_0^2/(16\beta_0)(12\beta_0 t+8\sin{(2\beta_0 t)}+\sin{(4\beta_0 t)})$.  
\begin{figure}[!ht]
\begin{center}
\includegraphics[width=0.9\linewidth]{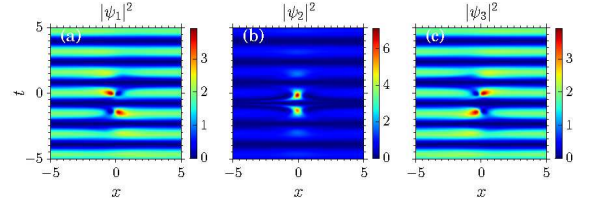}
\end{center}
\caption{(Color online) Contour plots: (a), (c) Four-petaled RW and (b) RW structure for $R(t)=1+\cos{(2 \beta_0 t)}$ and $\beta(t)^2=2\beta_0^2[1+3\tan^2(\beta_0 t)]$. The parameters are $A_1=1$, $A_2=5$, $A_3=0$, $A_4=0$,$r_0=1.0$, $\beta_0=2.5$, $b=0.01$, and $\delta=0.01$.}
\label{fig30}
\end{figure}
Here also we obtain the various localized structures when we vary the parameter $A_j$, $j=1,2,3,4$ and by fixing $\beta_0=0.1$.  The vector localized structures are similar to the previous two cases and so we do not display the outcome here separately.  The parameters are same as in Fig. \ref{fig18} with $\beta_0=2.5$.  In this case, the density profiles of $|\psi_1(x,t)|^2$, $|\psi_2(x,t)|^2$ and $|\psi_3(x,t)|^2$ show that single, two and three composite RWs exist on a periodic background as shown in Figs. \ref{fig30}-\ref{fig33}.    
\begin{figure}[!ht]
\begin{center}
\includegraphics[width=0.9\linewidth]{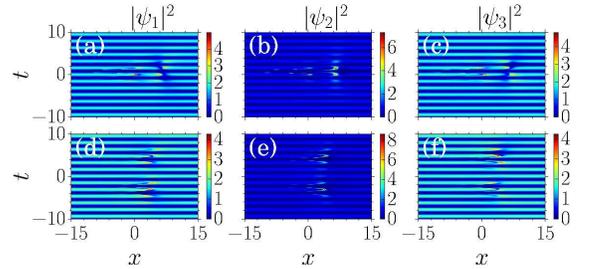} 
\end{center}
\caption{(Color online) Contour plots corresponding to two composite RWs: The parameters (a)-(c) $A_1=-2$, $A_2=30$, $A_3=6$ and $A_4=0$, and (d)-(f) $A_1=120$, $A_2=30$, $A_3=6$ and $A_4=0$.  The other parameters are same as in Fig. \ref{fig30}.}
\label{fig31}
\end{figure}
\begin{figure}[!ht]
\begin{center}
\includegraphics[width=0.9\linewidth]{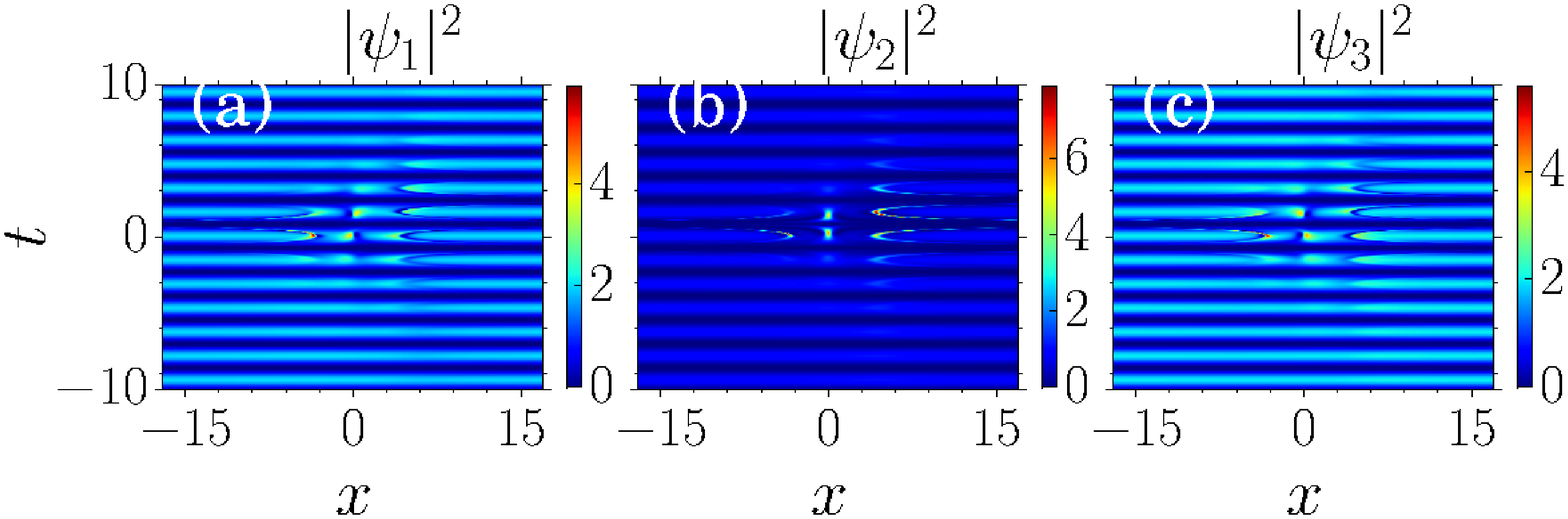} 
\end{center}
\caption{(Color online) Contour plots corresponding to three composite RWs: The parameters (a)-(c) $A_1=10$, $A_2=50$, $A_3=2$ and $A_4=5$. The other parameters are same as in Fig. \ref{fig30}.}
\label{fig32}
\end{figure}
\begin{figure}[!ht]
\begin{center}
\includegraphics[width=0.9\linewidth]{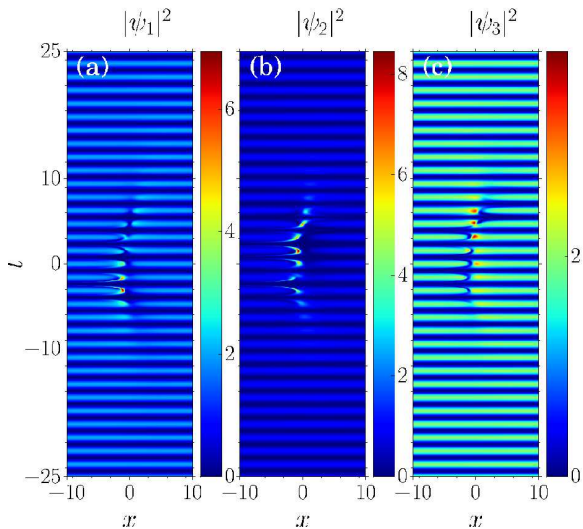} 
\end{center}
\caption{(Color online) Contour plots corresponding to the interactions of three composite RWs: The parameters (a)-(c) $A_1=150$, $A_2=250$, $A_3=120$ and $A_4=50$. The other parameters are same as in Fig. \ref{fig30}.}
\label{fig33}
\end{figure}
\section{Conclusion}
In conclusion, we have constructed several localized solutions of two coupled GPEs which describe the dynamics of the two-component BECs with  time-dependent scattering lengths and time-dependent harmonic trap potentials.  We have mapped the time-dependent two coupled GPEs onto two coupled NLSEs subject to an integrability condition between the time-dependent nonlinearity coefficient and the external trap potential through a similarity transformation technique. We have considered three different trap potentials, namely (i) time-independent trap, (ii) time-dependent monotonic trap and (iii) time-dependent periodic trap.  We have identified different localized wave structures, namely RWs, dark soliton-RW, bright soliton-RW and RW-breather-like structures for different values of particular parameters in the obtained solutions.  We have studied the characteristics of these localized solutions when we tune the strength of the trap parameter.  We have depicted the trajectories of nonautonomous RWs.  We have then constructed the dark-dark RW solutions for two-component BECs and investigated the correlated characteristics for the above trap potentials.  Our results show that in the case of time-independent trap potential, the RW structures maintain their stability by considering a lower value of trap parameter and then by increasing the trap parameter we find that the RW structures get more localized in time and stretched in space in a constant density background. In the case of time-dependent monotonic trap, the localized structures get compressed in time, delocalized in space and collapse also occurs when $t<0$ in each of the components due to the form of nature of potential.  In the case of periodic trap potential, RWs with soliton structures exist on a periodic background when we alter the trap parameter.  We have then constructed the general RW solutions for the mean-field model of three-component BECs. By restricting the parameters appearing in the obtained RW solutions, we have identified localized structures such as single, double and triple RWs for these trap potentials. We have analyzed how these localized density profiles get modified in the density background when we alter the trap parameter.  Our results show that in the case of time-independent trap potential the RW structures keep their stability when the trap parameter is small and the RWs get more localized in time and stretched in space when we increase the trap parameter. In the case of time-dependent monotonic trap, RW structures become compressed in time, delocalized in space and settled at a higher density background.  In the time-dependent periodic trap case, the RW structures emerge on a periodic background when we adjust the trap parameter.  Our results provide evidence for the existence of different possibilities to manipulate RWs experimentally in multi-component BEC systems.

\acknowledgments

K.M. thanks the University Grants Commission (UGC-RFSMS), Government of India, for providing support through a research fellowship.  The work of P.M. forms a part of Science and Engineering Research Board (SERB), Department of Science and Technology (DST), Govt. of India sponsored research project (No. EMR/2014/000644) and UGC-SAP (School of Physics, Bharathidasan University).  The work of M.S. forms part of a research project sponsored by National Board Higher Mathematics, Government of India. The work is also supported by a Department of Atomic Energy Raja Ramanna Fellowship for the project of M.L.

\end{document}